\documentstyle[12pt,epsf]{article}

\makeatletter
\def\textfraction{0.2}
\@textmin = \textfraction\textheight
\def\fps@figure{tbp} 
\def\fps@table{tbp} 
\makeatother

\renewcommand{\textfraction}{0}

\def\gsim{\ \rlap{\raise 3pt \hbox{$>$}}{\lower 3pt \hbox{$\sim$}}\ }
\def\lsim{\ \rlap{\raise 3pt \hbox{$<$}}{\lower 3pt \hbox{$\sim$}}\ }

\begin{document}

\begin{titlepage}

\begin{flushright}
CERN-TH/98-99\\
hep-ph/9805303
\end{flushright}
\vspace{0.2cm}

\begin{center}
\boldmath
\Large\bf
QCD Anatomy of $B\to X_s\gamma$ Decays
\unboldmath
\end{center}

\vspace{0.2cm}
\begin{center}
Alexander L. Kagan\\[0.1cm]
{\sl Department of Physics, University of Cincinnati\\
Cincinnati, Ohio 45221, USA}\\[0.3cm]
and\\[0.3cm]
Matthias Neubert\\[0.1cm]
{\sl Theory Division, CERN, CH-1211 Geneva 23, Switzerland}
\end{center}

\vspace{0.2cm}
\begin{abstract}
\vspace{0.2cm}\noindent
We present an updated next-to-leading order analysis of the $B\to
X_s\gamma$ branching ratio and photon spectrum, including consistently
the effects of Fermi motion in the heavy-quark expansion. For the
Standard Model, we obtain ${\rm B}(B\to X_s\gamma)=(2.57\pm
0.26_{-0.36}^{+0.31})\times 10^{-4}$ for the integral over the
high-energy part of the photon spectrum with $E_\gamma^{\rm lab}>
2.2$\,GeV, where the first error reflects the uncertainty in the input
parameters, and the second one the uncertainty in the calculation of
Fermi motion. This prediction agrees with the CLEO measurement of the
same quantity within one standard deviation. From a reanalysis of the
CLEO data, we obtain for the total branching ratio ${\rm B}(B\to
X_s\gamma)=(2.62\pm 0.60_{\rm exp}\,\mbox{}^{+0.37}_{-0.30\,{\rm
th}})\times 10^{-4}$ using the measured rate above 2.2\,GeV, and
$(2.66\pm 0.56_{\rm exp}\,\mbox{}^{+0.43}_{-0.48\,{\rm th}})\times
10^{-4}$ using a fit to the photon energy spectrum. Both values are
consistent with the Standard Model prediction of $(3.29\pm 0.33)\times
10^{-4}$. Our analysis contains an improved discussion of
renormalization scale dependence and QED corrections. We also study
the sensitivity of the branching ratio and photon spectrum to hadronic
parameters such as the $b$-quark mass, and to possible contributions
from New Physics beyond the Standard Model.
\end{abstract}

\vspace{0.5cm}
\centerline{(Submitted to European Physical Journal)}

\vfill
\noindent
CERN-TH/98-99\\
May 1998

\end{titlepage}

\section{Introduction}

The inclusive radiative decays $B\to X_s\gamma$ have been the subject
of a considerable number of experimental and theoretical
investigations. About three years ago, the CLEO Collaboration reported
a first measurement of the branching ratio for these decays, yielding
\cite{CLEO}
\begin{equation}
   \mbox{B}(B\to X_s\gamma)=(2.32\pm 0.57\pm 0.35)\times 10^{-4} \,,
\label{CLEObr}
\end{equation}
where the first error is statistical and the second is systematic
(including model dependence). Recently, the ALEPH Collaboration has
reported a measurement of the corresponding branching ratio for
$b$-hadrons produced at the $Z$ resonance, yielding \cite{ALEPH}
\begin{equation}
   \mbox{B}(H_b\to X_s\gamma)=(3.11\pm 0.80\pm 0.72)\times 10^{-4} \,,
\end{equation}
which is compatible with the CLEO result.

Being rare processes mediated by loop diagrams, inclusive radiative
decays are potentially sensitive probes of New Physics beyond the
Standard Model, provided a precise theoretical calculation of the
branching ratio can be performed. The general framework for such a
calculation is provided by the heavy-quark expansion, which predicts
that, up to small bound-state corrections, inclusive decay rates agree
with the parton model rates for the underlying decays of the $b$ quark
\cite{Chay}--\cite{MaWe}. As long as the fine structure of the photon
energy spectrum is not probed locally, the theoretical analysis of
$B\to X_s\gamma$ decays relies only on the weak assumption of global
quark--hadron duality. The leading nonperturbative corrections have
been studied in detail and are well understood
\cite{Adam}--\cite{Buch}. Still, the theoretical prediction for the
branching ratio suffers from large perturbative uncertainties of about
30\% if only leading-order expressions for the Wilson coefficient
functions in the effective weak Hamiltonian are employed
\cite{Gr90}--\cite{Poko}. Therefore, it was an important achievement
when last year the full next-to-leading order calculation of the total
$B\to X_s\gamma$ branching ratio in the Standard Model was completed,
combining consistently results for the matching conditions
\cite{Adel,GrHu}, matrix elements \cite{AliGr,Greub} and anomalous
dimensions \cite{Chet}. With this calculation the theoretical
uncertainty was reduced to a level of about 10\% \cite{Chet,Buras}.
More recently, the next-to-leading order analysis was also extended to
two-Higgs-doublet models \cite{Ciuc,Borzu}.

Whereas considerable effort has thus gone into calculating the total
$B\to X_s\gamma$ branching ratio, little progress has been made in
understanding the structure of the photon energy spectrum at
next-to-leading order. On the other hand, what is experimentally
accessible is only the high-energy part of the photon spectrum, and an
understanding of the spectral shape is thus a prerequisite for
extrapolating the data to the full phase space. For instance, the CLEO
Collaboration has measured the spectrum in the energy range between 2.2
and 2.7\,GeV (in the laboratory) and applied a correction factor of
$0.87\pm 0.06$ to extrapolate to the total decay rate
\cite{private}.\footnote{A similar treatment is followed in the ALEPH
analysis \protect\cite{ALEPH}.}
This factor does not take into account the full next-to-leading order
corrections to the decay rate. More importantly, it relies on an
estimate of bound-state effects \cite{Aliold} obtained using the
phenomenological model of Altarelli et al.\ \cite{ACM}, which is not
fully consistent with the heavy-quark expansion. The small uncertainty
assigned to the correction factor reflects the fact that the model
parameters (the Fermi momentum and the constituent quark masses) have
been tuned to fit the lepton spectrum in $B\to X_c\,\ell\,\nu$ decays
and then used to predict the photon spectrum in $B\to X_s\gamma$
decays. It is now known that there is no theoretical justification for
such a treatment \cite{shape}--\cite{shape2}. These conceptual
shortcomings were not improved in the updated analysis of the photon
spectrum presented by Ali and Greub a few years ago \cite{AliGr},
although more complete formulae for the perturbative corrections were
used in this work and a more conservative error analysis was presented.

The fact that only the high-energy part of the photon spectrum in $B\to
X_s\gamma$ decays is accessible experimentally introduces a significant
additional theoretical uncertainty \cite{adamtalks}, which has been
ignored in previous analyses. This observation limits the potential of
existing data on these decays to probe or constrain New Physics beyond
the Standard Model (for recent reviews, see Refs.~\cite{Gronau,Hewett}
and references therein). In this paper, we investigate in a systematic
way to what extent the high-energy part of the photon energy spectrum
in $B\to X_s\gamma$ decays can be controlled theoretically. The ``Fermi
motion'' of the $b$ quark inside the $B$ meson, which determines the
characteristic shape to the photon spectrum, can be consistently
described by taking a convolution of the parton model prediction for
the spectrum with a universal shape function $F(k_+)$, which determines
the light-cone momentum distribution of the $b$ quark in the $B$ meson
\cite{shape}--\cite{shape2}. We will for the first time present a
discussion of Fermi motion effects in a full next-to-leading order
analysis of $B\to X_s\gamma$ decays. In addition, our analysis contains
several improvements over previous works concerning, in particular, the
estimate of perturbative uncertainties, and the inclusion of QED
corrections. In Section~\ref{sec:rate}, we discuss in detail the
structure of the $B\to X_s\gamma$ branching ratio at next-to-leading
order in QCD, correcting some errors in the formulae for real-gluon
radiation contributions employed by previous authors. We introduce a
decomposition of the branching ratio in terms of the values of the
Wilson coefficients $C_i(m_W)$ at the weak scale, which is particularly
convenient to discuss the sensitivity to New Physics beyond the
Standard Model. This decomposition is also the starting point of a
thorough discussion of the renormalization-scale dependence. We find
that the perturbative uncertainty in the theoretical prediction for the
branching ratio has been underestimated by previous authors
\cite{Chet}--\cite{Borzu} by more than a factor of 2. We also suggest a
new definition of the ``total'' branching ratio, which is insensitive
to the unphysical soft-photon divergence in the $b\to sg\gamma$
subprocess. In Section~\ref{sec:Fermi}, we show that Fermi motion
effects, which result from the residual interaction of the $b$ quark
inside the $B$ meson, give rise to the dominant theoretical uncertainty
in the calculation of the partially integrated (over photon energy)
branching ratio. We present a consistent treatment of these effects
based on first principles of the heavy-quark expansion, emphasizing
that for such partially integrated quantities the main element of
uncertainty in the description of Fermi motion lies in the value of the
$b$-quark mass. Other features associated with the detailed functional
form of the shape function play a minor role. We make a prediction
for the $B\to X_s\gamma$ branching ratio with a restriction on the
photon energy such that $E_\gamma^{\rm lab}>2.2$\,GeV and find
agreement with the CLEO measurement of the same quantity within one
standard deviation. We also extract a new value for the total branching
ratio, which is significantly different from the published CLEO result
reported in (\ref{CLEObr}). In Section~\ref{sec:spectrum}, we extend
the discussion to the photon spectrum itself, investigating first the
structure of the contributions from different operators in the
effective Hamiltonian. We find that, to a high degree of accuracy, the
shape of the photon spectrum is determined by QCD dynamics and is
insensitive to New Physics beyond the Standard Model. We then perform a
fit of our theoretical predictions to the CLEO data on the photon
spectrum and extract again a value for the total branching ratio. We
also discuss the possibility of determining, from future high-precision
data on the photon spectrum, a value of the $b$-quark mass with a
well-defined short-distance interpretation. In Section~\ref{sec:mass},
we consider the hadronic invariant mass spectrum and discuss the role
of quark--hadron duality in the comparison of experimental data with
our theoretical predictions. We derive a realistic, one-parameter
description of the spectrum that is valid even in the low-mass region,
where quark--hadron duality breaks down. Finally, in
Section~\ref{sec:NP} we explore how New Physics beyond the Standard
Model may affect the spectral shape and the total branching ratio in
$B\to X_s\gamma$ decays. Section~\ref{sec:concl} contains the
conclusions. The paper also comprises three Appendices, where we
discuss QED corrections to the $B\to X_s\gamma$ branching ratio, the
Doppler broadening of the photon spectrum in the decays of $B$ mesons
produced at the $\Upsilon(4s)$ resonance, and technical details of the
calculation of the photon energy spectrum.

\boldmath
\section{$B\to X_s\gamma$ branching ratio}
\unboldmath
\label{sec:rate}

The theoretical analysis of the $B\to X_s\gamma$ branching ratio at
next-to-leading order has been discussed previously by several authors.
In this section we review the main ingredients of the calculation. In
addition, we present several improvements of it, which concern the
treatment of leading-logarithmic QED corrections, the analysis of the
renormalization-scale dependence, and a discussion of the sensitivity
to New Physics. We also correct some mistakes in the results for
real-gluon emission presented in the literature.

The starting point in the calculation of inclusive $B$ decay rates is
the low-energy effective Hamiltonian \cite{Heff}
\begin{equation}
   H_{\rm eff} = -\frac{4 G_F}{\sqrt2}\,V_{ts}^* V_{tb}
   \sum_i C_i(\mu_b) O_i(\mu_b) \,.
\label{Heff}
\end{equation}
The operators relevant to our discussion are
\begin{eqnarray}
   O_2 &=& \bar s_L\gamma_\mu c_L\bar c_L\gamma^\mu b_L \,, \nonumber\\
   O_7 &=& \frac{e\,m_b}{16\pi^2}\,\bar s_L\sigma_{\mu\nu}
    F^{\mu\nu} b_R \,, \nonumber\\
   O_8 &=& \frac{g_s m_b}{16\pi^2}\,\bar s_L\sigma_{\mu\nu}
    G_a^{\mu\nu} t_a b_R \,.
\end{eqnarray}
To an excellent approximation, the contributions of other operators can
be neglected. The renormalization scale $\mu_b$ in (\ref{Heff}) is
conveniently chosen of order $m_b$, so that all large logarithms reside
in the Wilson coefficient functions. The complete theoretical
prediction for the $B\to X_s\gamma$ decay rate at next-to-leading order
has been presented for the first time by Chetyrkin et al.\ \cite{Chet}.
It depends on a parameter $\delta$ defined by the condition that the
photon energy be above a threshold given by $E_\gamma>(1-\delta)
E_\gamma^{\rm max}$, where $E_\gamma^{\rm max}=m_b/2$ is the maximum
photon energy attainable in the parton model. (Throughout this paper,
we will neglect the mass of the strange quark whenever possible.) The
prediction for the $B\to X_s\gamma$ branching ratio is usually obtained
by normalizing the result for the corresponding decay rate to that for
the semileptonic decay rate, thereby eliminating a strong dependence on
the $b$-quark mass. We define
\begin{equation}
   R_{\rm th}(\delta) =
   \frac{\Gamma(B\to X_s\gamma)\big|_{E_\gamma>(1-\delta)
         E_\gamma^{\rm max}}}{\Gamma(B\to X_c\,e\,\bar\nu)}
   = \frac{6\alpha}{\pi f(z)}\,\left| \frac{V_{ts}^* V_{tb}}{V_{cb}}
    \right|^2 K_{\rm NLO}(\delta) \,,
\label{GNLO}
\end{equation}
where $f(z)=1-8z+8z^3-z^4-12z^2\ln z\approx 0.542-2.23(\sqrt z-0.29)$
is a phase-space factor depending on the mass ratio $z=(m_c/m_b)^2$,
for which we shall take $\sqrt z=0.29\pm 0.02$. In the context of our
analysis, the quark masses are defined as one-loop pole masses. The
electro-magnetic coupling $\alpha=1/137.036$ is the fine-structure
constant renormalized at $q^2=0$, as is appropriate for real-photon
emission \cite{CzMa}. The quantity $K_{\rm NLO}(\delta)=|C_7|^2+\dots$
contains the corrections to the leading-order result. In terms of the
theoretically calculable ratio $R_{\rm th}(\delta)$, the $B\to
X_s\gamma$ branching ratio is given by
\begin{equation}
   {\rm B}(B\to X_s\gamma)\big|_{E_\gamma>(1-\delta)E_\gamma^{\rm max}}
   = R_{\rm th}(\delta)\times {\rm B}(B\to X_c\,e\,\bar\nu)
   = 0.105 N_{\rm SL}\,R_{\rm th}(\delta) \,,
\label{numeric}
\end{equation}
where $N_{\rm SL}={\rm B}(B\to X_c\,e\,\bar\nu)/10.5\%$ is a
normalization factor to be determined from experiment. To good
approximation $N_{\rm SL}=1$. The current experimental situation of
measurements of the semileptonic branching ratio of $B$ mesons and
their theoretical interpretation are reviewed in
Refs.~\cite{Persis,me}.

In the calculation of the quantity $K_{\rm NLO}(\delta)$ we shall
consistently work to first order in the small parameters $\alpha_s$,
$1/m_Q^2$ and $\alpha/\alpha_s$, the latter ratio being related to the
leading-logarithmic QED corrections. The general structure of the
result is
\begin{eqnarray}
   K_{\rm NLO}(\delta) &=& \sum_{ \stackrel{i,j=2,7,8}{i\le j} }
    k_{ij}(\delta,\mu_b)\,\mbox{Re}\!\left[ C_i^{(0)}(\mu_b)\,
    C_j^{(0)*}(\mu_b) \right]
    + S(\delta)\,\frac{\alpha_s(\mu_b)}{2\pi}\,
    \mbox{Re}\!\left[ C_7^{(1)}(\mu_b)\,C_7^{(0)*}(\mu_b) \right]
    \nonumber\\
   &&\mbox{}+ S(\delta)\,\frac{\alpha}{\alpha_s(\mu_b)} \bigg(
    2\,\mbox{Re}\!\left[ C_7^{({\rm em})}(\mu_b)\,C_7^{(0)*}(\mu_b)
    \right] - k_{\rm SL}^{({\rm em})}(\mu_b)\,|C_7^{(0)}(\mu_b)|^2
    \bigg) \,,
\label{KNLO}
\end{eqnarray}
where we have expanded the Wilson coefficients as\footnote{These are
the effective, scheme-independent Wilson coefficient functions
introduced in Ref.~\protect\cite{Poko}.}
\begin{equation}
   C_i(\mu_b) = C_i^{(0)}(\mu_b) + \frac{\alpha_s(\mu_b)}{4\pi}\,
   C_i^{(1)}(\mu_b) + \frac{\alpha}{\alpha_s(\mu_b)}\,
   C_i^{({\rm em})}(\mu_b) + \dots \,.
\label{Ciexp}
\end{equation}
The leading-order coefficients are given by
\begin{eqnarray}
   C_2^{(0)}(\mu_b) &=& \frac 12 \left( \eta^{-\frac{12}{23}}
    + \eta^\frac{6}{23} \right) \,, \nonumber\\
   C_7^{(0)}(\mu_b) &=& \eta^\frac{16}{23}\,C_7^{(0)}(m_W)
    + \frac 83 \left( \eta^\frac{14}{23} - \eta^\frac{16}{23} \right)
    C_8^{(0)}(m_W) + \sum_{i=1}^8\,h_i\,\eta^{a_i} \,, \nonumber\\
   C_8^{(0)}(\mu_b) &=& \eta^\frac{14}{23}\,C_8^{(0)}(m_W)
    + \sum_{i=1}^8\,\bar h_i\,\eta^{a_i} \,,
\label{evol}
\end{eqnarray}
where $\eta=\alpha_s(m_W)/\alpha_s(\mu_b)$, and $h_i$, $\bar h_i$ and
$a_i$ are known numerical coefficients \cite{Guidoetal,Poko}. In the
Standard Model, the Wilson coefficients of the dipole operators at the
scale $m_W$ are functions of the mass ratio
$x_t=(\overline{m}_t(m_W)/m_W)^2$ given by \cite{Gr90}
\begin{eqnarray}
   C_7^{(0)}(m_W) &=& \frac{3x_t^3 - 2x_t^2}{4(x_t-1)^4}
    \ln x_t + \frac{-8x_t^3 - 5x_t^2 + 7x_t}{24(x_t-1)^3} \,,
    \nonumber\\
   C_8^{(0)}(m_W) &=& \frac{-3x_t^2}{4(x_t-1)^4} \ln x_t
    + \frac{-x_t^3 + 5x_t^2 + 2x_t}{8(x_t-1)^3} \,.
\label{GrSpr}
\end{eqnarray}
The next-to-leading terms in (\ref{Ciexp}) must be kept only for the
coefficient $C_7(\mu_b)$. The expression for $C_7^{(1)}(\mu_b)$ can be
found in eq.~(21) of Ref.~\cite{Chet}. Our treatment of QED corrections
differs from that of Czarnecki and Marciano \cite{CzMa} in that we
perform a renormalization-group improvement to resum the contributions
to $C_7(\mu_b)$ of order $\alpha L\,(\alpha_s L)^n$, with
$L=\ln(m_W/\mu_b)$, to all orders in perturbation theory, whereas these
authors include only the terms with $n=0$. Numerically, the resummation
decreases the effect of QED correction by almost a factor of 2. The
technical details of our calculation are discussed in Appendix~A. The
result for $C_7^{({\rm em})}(\mu_b)$ is
\begin{eqnarray}
   C_7^{({\rm em})}(\mu_b) &=&
    \left( \frac{32}{75}\,\eta^{-\frac{9}{23}}
    - \frac{40}{69}\,\eta^{-\frac{7}{23}}
    + \frac{88}{575}\,\eta^{\frac{16}{23}} \right) C_7^{(0)}(m_W)
    \nonumber\\
   &&\mbox{}+
    \left( -\frac{32}{575}\,\eta^{-\frac{9}{23}}
    + \frac{32}{1449}\,\eta^{-\frac{7}{23}}
    + \frac{640}{1449}\,\eta^{\frac{14}{23}}
    - \frac{704}{1725}\,\eta^{\frac{16}{23}} \right) C_8^{(0)}(m_W)
    \nonumber\\
   &&\mbox{}-\frac{190}{8073}\,\eta^{-\frac{35}{23}}
    - \frac{359}{3105}\,\eta^{-\frac{17}{23}}
    + \frac{4276}{121095}\,\eta^{-\frac{12}{23}}
    + \frac{350531}{1009125}\,\eta^{-\frac{9}{23}} \nonumber\\
   &&\mbox{}+ \frac{2}{4347}\,\eta^{-\frac{7}{23}}
    - \frac{5956}{15525}\,\eta^{\frac{6}{23}}
    + \frac{38380}{169533}\,\eta^{\frac{14}{23}}
    - \frac{748}{8625}\,\eta^{\frac{16}{23}} \,.
   \nonumber\\
\label{fancy}
\end{eqnarray}
The result for the leading QED correction to the semileptonic decay
rate is \cite{Sirl}
\begin{equation}
   k_{\rm SL}^{({\rm em})}(\mu_b)
   = \frac{12}{23} \left( \eta^{-1} - 1 \right)
   = \frac{2\alpha_s(\mu_b)}{\pi} \ln\frac{m_W}{\mu_b} \,.
\label{kSL}
\end{equation}
We note that unlike the factor $\alpha$ in (\ref{GNLO}), which results
from the calculation of a matrix element for a process with real-photon
emission, the QED corrections to the Wilson coefficients arise from the
evolution of local operators, and hence the coupling $\alpha$ in
(\ref{Ciexp}) should in principle be taken as a running coupling
$\alpha(\mu)$ rather than the fine-structure constant renormalized at
$q^2=0$. However, including the running of the QED coupling in the
operator evolution would only induce corrections of order $(\alpha
L)^2(\alpha_s L)^n$, which from a numerical point of view can be safely
neglected.

For the purpose of illustration, we note that with $\mu_b=4.8$\,GeV the
values of the various coefficients in the Standard Model are:
$C_2^{(0)}(m_b)\approx 1.11$, $C_7^{(0)}(m_b)\approx -0.31$,
$C_8^{(0)}(m_b)\approx -0.15$, as well as $C_7^{(1)}(m_b)\approx 0.48$
and $C_7^{({\rm em})}(m_b)\approx 0.03$. The QED correction
proportional to $C_7^{({\rm em})}$ in (\ref{KNLO}) is about a factor
0.13 smaller than the next-to-leading order QCD correction proportional
to $C_7^{(1)}$.

The coefficient functions $k_{ij}(\delta,\mu_b)$ in (\ref{KNLO}) are
given by
\begin{eqnarray}
   k_{77}(\delta,\mu_b) &=& S(\delta) \left\{ 1
    + \frac{\alpha_s(\mu_b)}{2\pi} \left( r_7
    + \gamma_{77}\ln\frac{m_b}{\mu_b} - \frac{16}{3} \right)
    + \left[ \frac{(1-z)^4}{f(z)} - 1\right]
    \frac{6\lambda_2}{m_b^2} \right\} \nonumber\\
   &&\mbox{}+ \frac{\alpha_s(\mu_b)}{\pi}\,f_{77}(\delta)
    + S(\delta)\,\frac{\alpha_s(\bar\mu_b)}{2\pi}\,\bar\kappa(z) \,,
    \nonumber\\
   k_{27}(\delta,\mu_b) &=& S(\delta) \left[
    \frac{\alpha_s(\mu_b)}{2\pi}
    \left( \mbox{Re}(r_2) + \gamma_{27}\ln\frac{m_b}{\mu_b} \right)
    - \frac{\lambda_2}{9 m_c^2} \right]
    + \frac{\alpha_s(\mu_b)}{\pi}\,f_{27}(\delta) \,, \nonumber\\
   k_{78}(\delta,\mu_b) &=& S(\delta)\,\frac{\alpha_s(\mu_b)}{2\pi}
    \left( \mbox{Re}(r_8) + \gamma_{87}\ln\frac{m_b}{\mu_b} \right)
    + \frac{\alpha_s(\mu_b)}{\pi}\,f_{78}(\delta) \,, \nonumber\\
   k_{ij}(\delta,\mu_b) &=& \frac{\alpha_s(\mu_b)}{\pi}\,
    f_{ij}(\delta) \,;\qquad \{i,j\}=\{2,2\},\,\{8,8\},\,\{2,8\} \,,
\label{kij}
\end{eqnarray}
where
\begin{equation}
   S(\delta) = \exp\!\left[ -\frac{2\alpha_s(\mu_b)}{3\pi}
   \left( \ln^2\!\delta + \frac 72\ln\delta \right) \right]
\label{Suda}
\end{equation}
is a Sudakov factor, $\gamma_{77}=\frac{32}{3}$, $\gamma_{27}=
\frac{416}{81}$ and $\gamma_{87}=-\frac{32}{9}$ are entries of the
anomalous dimension matrix, and
\begin{eqnarray}
   r_7 &=& -\frac{10}{3} - \frac{8\pi^2}{9} \,, \qquad
    \mbox{Re}(r_8) = \frac{44}{9} - \frac{8\pi^2}{27} \,, \nonumber\\
   \mbox{Re}(r_2) &\approx& -4.092 + 12.78(\sqrt z-0.29)
\end{eqnarray}
are numerical coefficients resulting from the calculation of the matrix
elements of the local operators $O_i$ in the effective Hamiltonian at
next-to-leading order \cite{Greub}. Finally, $\bar\kappa(z)\approx
3.382-4.14(\sqrt z-0.29)$ is the next-to-leading correction to the
semileptonic decay rate \cite{Nir89}. To $O(\alpha_s)$ the explicit
$\mu_b$ dependence of the coefficients $k_{ij}(\delta,\mu_b)$ cancels
against that of the Wilson coefficient functions. Following
Ref.~\cite{Buras}, we allow for different renormalization scales in
radiative and semileptonic $B$ decays (i.e.\ $\mu_b\ne\bar\mu_b$).

The functions $f_{ij}(\delta)$ in (\ref{kij}) account for the effects
of real-gluon radiation and are defined such that $f_{ij}(0)=0$. They
can be obtained from results given in Refs.~\cite{AliGr,Greub} by
performing some phase-space integrations. We find
\begin{eqnarray}
   f_{77}(\delta) &=& \frac{1}{3}\,\bigg[ 10\delta + \delta^2
    - \frac{2\delta^3}{3} + \delta(\delta-4)\ln\delta \bigg] \,,
    \nonumber\\
   f_{88}(\delta) &=& \frac{1}{27}\,\Bigg\{ 4 L_2(1-\delta)
    - \frac{2\pi^2}{3} + 8\ln(1-\delta) - \delta(2+\delta)\ln\delta
    \nonumber\\
   &&\hspace{0.7cm}\mbox{}+ 7\delta + 3\delta^2 - \frac{2\delta^3}{3}
    - 2 \left[ 2\delta + \delta^2 + 4\ln(1-\delta) \right]
    \ln\frac{m_b}{m_s} \Bigg\} \,, \nonumber\\
   f_{78}(\delta) &=& \frac{8}{9}\,\Bigg[ L_2(1-\delta)
    - \frac{\pi^2}{6} - \delta\ln\delta + \frac{9\delta}{4}
    - \frac{\delta^2}{4} + \frac{\delta^3}{12} \Bigg] \,, \nonumber\\
   f_{22}(\delta) &=& \frac{16}{27} \int\limits_0^1\!{\rm d}x\,
    (1-x)(1-x_\delta)\,\left|\, \frac{z}{x}\,
    G\!\left(\frac{x}{z}\right) + \frac 12 \,\right|^2 \,, \nonumber\\
   f_{27}(\delta) &=& -3 f_{28}(\delta)
    = - \frac{8z}{9} \int\limits_0^1\!{\rm d}x\,(1-x_\delta)\,
    \mbox{Re}\!\left[ G\!\left(\frac{x}{z}\right) + \frac{x}{2z}
    \right] \,,
\label{fij}
\end{eqnarray}
where $x_\delta=\mbox{max}(x,1-\delta)$, and
\begin{equation}
   G(t) = \left\{ \begin{array}{cl}
    -2\arctan^2\!\sqrt{t/(4-t)} & ~;~t<4 \,, \\[0.1cm]
    2 \left( \ln\Big[(\sqrt{t}+\sqrt{t-4})/2\Big]
    - \displaystyle\frac{i\pi}{2} \right)^2 & ~;~t\ge 4 \,.
   \end{array} \right.
\end{equation}
Our expressions for $f_{78}(\delta)$ and $f_{88}(\delta)$ disagree with
the corresponding ones in Ref.~\cite{Chet}, which have later been used
by several authors. (The corrected expressions are also given in an
Erratum to Ref.~\protect\cite{Chet}.) We shall comment on the numerical
effect of this correction below. The function $f_{88}$ is sensitive to
collinear singularities regulated by the mass of the strange quark. The
collinear logarithms can be resummed to all orders of perturbation
theory, leading to a collinear-safe result \cite{Lige}. Unless $\delta$
is chosen very close to 1, the net effect of the resummation is a
moderate increase of the result. Since the contribution proportional to
$f_{88}$ is very small, however, it is sufficient for all practical
purposes to work with the leading-order expression given above. We take
a rather large value for the quark-mass ratio, $m_b/m_s=50$, in order
to mimic the effect of the resummation of collinear logarithms.

Bound-state corrections enter the theoretical expressions for the
coefficients $k_{ij}$ at order $1/m_Q^2$ and are proportional to the
hadronic parameter $\lambda_2=\frac 14(m_{B^*}^2-m_B^2)\approx
0.12$\,GeV$^2$ \cite{FaNe}. The corrections proportional to $1/m_b^2$
entering the expression for $k_{77}$ characterize a spin-dependent
interaction inside the $B$ meson \cite{Bigi}--\cite{Adam}. A peculiar
feature of inclusive radiative decays is the appearance of the
correction proportional to $1/m_c^2$ in $k_{27}$, which represents a
long-distance contribution arising from $(c\bar c)$ intermediate states
\cite{Volo,Khod}. Strictly speaking, these effects are non-local in
nature; however, to a good approximation they can be represented by a
local $1/m_c^2$ correction \cite{LRW97,Grant}. The correct sign of this
term has only recently been found in Ref.~\cite{Buch}.

Finally, a comment is in order about our treatment of the Sudakov
factor, which is slightly different from that of previous authors. In
Refs.~\cite{Chet}--\cite{Ciuc}, the Sudakov factor was only included
for the leading term in $k_{77}$. If this is done, the decay rate
becomes negative for small values of $\delta$, which is an unphysical
result. All terms not vanishing in the limit $\delta=0$ correspond to a
two-body decay $b\to s\gamma$ and must be suppressed by a Sudakov
factor. Once this is done, the quantity $K_{\rm NLO}(\delta)$ vanishes
in the limit $\delta\to 0$ as it should. A full next-to-leading order
resummation of Sudakov logarithms that goes beyond the naive
exponentiation of the one-loop result shown in (\ref{Suda}) is possible
but rather complicated. In Refs.~\cite{Korch,Akhou}, such a resummation
has been performed for high-order moments of the photon energy
spectrum; however, the results are such that a numerical evaluation
would require integration over the running coupling constant
$\alpha_s(k_\perp)$ in the region $\Lambda_{\rm QCD}<k_\perp<m_b$. In
Ref.~\cite{MN}, the resummation has been extended to the partially
integrated photon spectrum itself (rather than its moments), and a
factorization of short- and long-distance contributions has been
performed such that all contributions from scales $k_\perp$ in the
range $(\Lambda_{\rm QCD} m_b)^{1/2}<k_\perp<m_b$ are treated
perturbatively, whereas contributions from scales in the range
$\Lambda_{\rm QCD}<k_\perp< (\Lambda_{\rm QCD} m_b)^{1/2}$ are absorbed
into the definition of the shape function. This guarantees that the
resummed formulae can be reliably evaluated in perturbation theory. If
this is done, it turns out that the resummation is a very small effect,
which can be neglected for all practical purposes \cite{MN}.

In order to explore the sensitivity of the theoretical prediction for
the $B\to X_s\gamma$ branching ratio to possible New Physics
contributions, it is instructive to make explicit the dependence of the
result on the values of the Wilson coefficients of the dipole operators
$O_7$ and $O_8$ at the scale $m_W$. To this end, we introduce the
ratios
\begin{equation}
   \xi_7 = \frac{C_7(m_W)}{C_7^{\rm SM}(m_W)} \,, \qquad
   \xi_8 = \frac{C_8(m_W)}{C_8^{\rm SM}(m_W)} \,.
\label{xiLdef}
\end{equation}
They are normalized to the Standard Model contributions, which at
next-to-leading order take the values $C_7^{\rm SM}(m_W)\approx-0.22$
and $C_8^{\rm SM}(m_W)\approx-0.12$ \cite{Adel,GrHu}. In many
extensions of the Standard Model there are contributions to $C_7(m_W)$
and $C_8(m_W)$ from new flavour physics at a high scale, and
consequently the parameters $\xi_7$ and $\xi_8$ may take (even complex)
values different from 1. Similarly, New Physics may induce dipole
operators with opposite chirality to that of the Standard Model, i.e.\
operators with right-handed light-quark fields. If we denote by $C_7^R$
and $C_8^R$ the Wilson coefficients of these new operators, the
expression (\ref{KNLO}) can be modified to include their contributions
by simply replacing $C_i C_j^*\to C_i C_j^* + C_i^R C_j^{R*}$
everywhere, taking however into account that $C_2^R=0$. We thus define
two additional parameters
\begin{equation}
   \xi_7^R = \frac{C_7^R(m_W)}{C_7^{\rm SM}(m_W)} \,, \qquad
   \xi_8^R = \frac{C_8^R(m_W)}{C_8^{\rm SM}(m_W)} \,,
\label{xiRdef}
\end{equation}
which vanish in the Standard Model. Since the dipole operators
only contribute to rare flavour-changing neutral current processes,
there are at present rather weak constraints on the values of these
parameters (see Section~\ref{sec:NP}). On the other hand, we assume
that the coefficient $C_2$ of the current--current operator $O_2$ takes
its Standard Model value, and that there is no similar operator
containing right-handed quark fields. Since the operator $O_2$ mediates
Cabibbo-allowed decays of $B$ mesons, any significant New Physics
contribution to $C_2$ would already have been detected experimentally.

With these definitions, the $B\to X_s\gamma$ branching ratio can be
decomposed as
\begin{eqnarray}
   &&\hspace{-0.5cm}
    \frac{1}{N_{\rm SL}}\,{\rm B}(B\to X_s\gamma)
    \big|_{E_\gamma>(1-\delta)E_\gamma^{\rm max}} \nonumber\\
   &=& B_{22}(\delta) + B_{77}(\delta) \left( |\xi_7|^2 + |\xi_7^R|^2
    \right) + B_{88}(\delta) \left( |\xi_8|^2 + |\xi_8^R|^2 \right)
    \nonumber\\
   &&\mbox{}+ B_{27}(\delta)\,\mbox{Re}(\xi_7) + B_{28}(\delta)\,
    \mbox{Re}(\xi_8) + B_{78}(\delta) \left[ \mbox{Re}(\xi_7\xi_8^*)
    + \mbox{Re}(\xi_7^R\xi_8^{R*}) \right] \,.
\label{xidecomp}
\end{eqnarray}
The components $B_{ij}(\delta)$ are formally independent of the
renormalization scale $\mu_b$. Their residual scale dependence results
only from the truncation of perturbation theory at next-to-leading
order. In Table~\ref{tab:kij_parton}, the values of these quantities
are given for different choices of the renormalization scale and the
cutoff on the photon energy. The input parameters entering the
calculation will be discussed below. Typically, the components $B_{ij}$
vary by amounts of order 10--20\% as $\mu_b$ varies between $m_b/2$ and
$2m_b$. The good stability is a result of the explicit cancelation of
the $\mu_b$ dependence between the Wilson coefficients and matrix
elements achieved by a full next-to-leading order calculation. The
Standard Model branching ratio is obtained by adding the various
contributions setting $\xi_7=\xi_8=1$ and $\xi_7^R=\xi_8^R=0$, as shown
in the last column. The most important contributions are the 2-2 and
2-7 terms, followed by the 7-7 term. Note that with a realistic choice
of the cutoff parameter $\delta$ the coefficient $B_{88}(\delta)$ of
the term proportional to $|\xi_8|^2+|\xi_8^R|^2$ is very small.
Therefore, $B\to X_s\gamma$ decays have a low sensitivity to enhanced
chromo-magnetic dipole transitions. For the remainder of this section
we focus on the Standard Model and evaluate the various theoretical
uncertainties in the prediction for the branching ratio. The impact of
New Physics will be discussed in Section~\ref{sec:NP}.

\begin{table}
\centerline{\parbox{14cm}{\caption{\label{tab:kij_parton}\small\sl
Values of the coefficients $B_{ij}(\delta)$ in units of $10^{-4}$, for
different choices of $\mu_b$}}}
\begin{center}
\begin{tabular}{|cc|cccccc|c|}
\hline
$\mu_b$ & $\delta$ & $B_{22}$ & $B_{77}$ & $B_{88}$ & $B_{27}$
 & $B_{28}$ & $B_{78}$ & $\sum B_{ij}$ \\
\hline\hline
$m_b/2$
 & 0.90 & 1.321 & 0.335 & 0.015 & 1.265 & 0.179 & 0.074 & 3.188 \\
 & 0.30 & 1.167 & 0.322 & 0.005 & 1.196 & 0.136 & 0.070 & 2.896 \\
 & 0.15 & 1.080 & 0.309 & 0.004 & 1.143 & 0.126 & 0.067 & 2.728 \\
\hline
$m_b$
 & 0.90 & 1.258 & 0.382 & 0.015 & 1.395 & 0.161 & 0.083 & 3.293 \\
 & 0.30 & 1.239 & 0.361 & 0.005 & 1.387 & 0.137 & 0.080 & 3.210 \\
 & 0.15 & 1.200 & 0.347 & 0.004 & 1.354 & 0.132 & 0.077 & 3.114 \\
\hline
$2 m_b$
 & 0.90 & 1.023 & 0.428 & 0.015 & 1.517 & 0.132 & 0.092 & 3.206 \\
 & 0.30 & 1.041 & 0.402 & 0.004 & 1.552 & 0.118 & 0.091 & 3.209 \\
 & 0.15 & 1.021 & 0.386 & 0.004 & 1.535 & 0.115 & 0.088 & 3.150 \\
\hline
\end{tabular}
\end{center}
\end{table}

\begin{figure}
\epsfxsize=8cm
\centerline{\epsffile{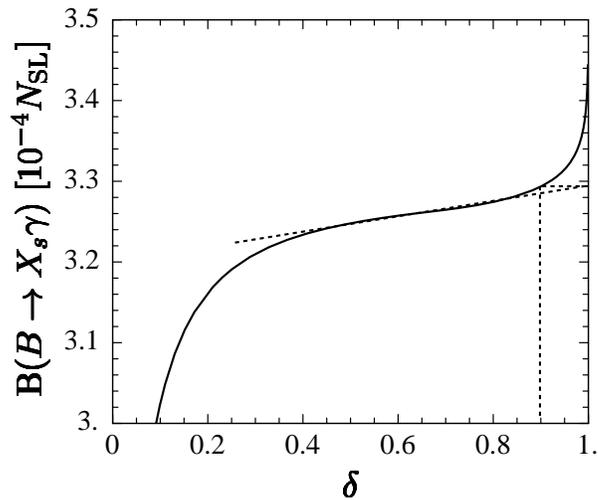}}
\centerline{\parbox{14.5cm}{\caption{\label{fig:delta}\small\sl
Dependence of the $B\to X_s\gamma$ branching ratio on the cutoff
parameter $\delta$. The dashed lines indicate the extrapolation to the
``total'' branching ratio.}}}
\end{figure}

In Table~\ref{tab:kij_parton}, the choice $\delta=0.9$ corresponds to
the unrealistic case of an almost fully inclusive measurement, whereas
$\delta=0.3$ and 0.15 correspond to restrictions to the high-energy
part of the photon spectrum, which in practice is required for
experimental reasons. The theoretical prediction for the branching
ratio diverges in the limit $\delta\to 1$ because of a logarithmic
singularity in the term proportional to $f_{88}(\delta)$, which
reflects the soft-photon divergence of the $b\to sg\gamma$ subprocess.
Previous authors \cite{Chet}--\cite{Ciuc} have chosen to define the
``total'' $B\to X_s\gamma$ branching ratio by taking $\delta=0.99$. In
our opinion this is not the best definition possible, because for
values of $\delta$ so close to 1 the theoretical result becomes very
sensitive to the unphysical soft-photon divergence. This is evident
from Figure~\ref{fig:delta}, which shows the integrated branching ratio
as a function of $\delta$. We believe a more reasonable definition of
the ``total'' branching ratio is to use an extrapolation to $\delta=1$
starting from the region $\delta\sim 0.5$--0.8, where the theoretical
result exhibits a weak, almost linear dependence on the cutoff. The
simple geometric construction indicated by the dashed lines in the
figure shows that the extrapolated value so defined agrees, to a good
approximation, with the result obtained by taking $\delta=0.9$. Hence,
from now on we define the ``total'' branching ratio to be that
corresponding to this particular value of the cutoff.

\begin{figure}
\epsfxsize=14cm
\centerline{\epsffile{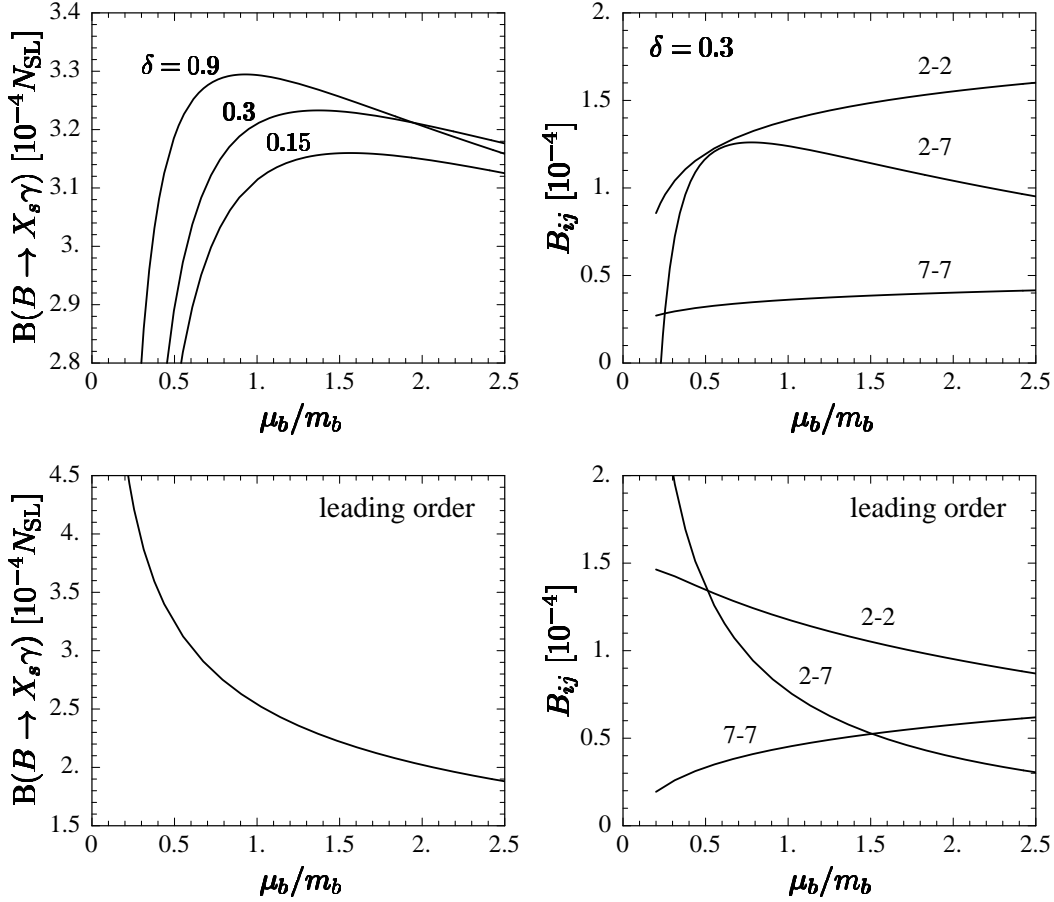}}
\centerline{\parbox{14.5cm}{\caption{\label{fig:mub_dep}\small\sl
Scale dependence of the $B\to X_s\gamma$ branching ratio (left) and of
its three most important components (right) in the Standard Model. For
comparison, the lower plots show the results obtained at leading
order.}}}
\end{figure}

The dependence of the theoretical results on the choice of the
renormalization scale is conventionally taken as an estimate of
higher-order corrections. Following common practice, we vary the
renormalization scales $\mu_b$ and $\bar\mu_b$ independently in the
range between $m_b/2$ and $2 m_b$; their central values are taken to be
$m_b$. The dependence on the scale $\bar\mu_b$ entering the formula for
the semileptonic decay rate is straightforward to analyse. The result
is shown in Table~\ref{tab:errors}. The analysis of the scale
dependence of the radiative decay rate is more subtle. Previous authors
have estimated the $\mu_b$ dependence of the total branching ratio in
the Standard Model and found a striking improvement over the leading
order result. A dedicated discussion of this issue has been presented
by Buras et al.\ \cite{Buras}, and the results obtained in this work
were later confirmed in Refs.~\cite{Ciuc,Borzu}. One finds a variation
of the total branching ratio by $\mbox{}^{+0.1}_{-3.2}\%$, as compared
with $\mbox{}^{+27.4}_{-20.4}\%$ at leading order. We agree with those
results, but we believe they cannot be taken as a realistic estimate of
the size of unknown higher-order corrections. The excellent stability
observed at next-to-leading order is largely due to an accidental
cancelation between different contributions to the decay rate. This
point of view is supported by the fact that in some extensions of the
Standard Model, such as two-Higgs-doublet models, a much stronger scale
dependence is observed in some regions of parameter space \cite{Borzu}.
On the left-hand side in Figure~\ref{fig:mub_dep}, we show the
branching ratio as a function of $\mu_b/m_b$, both at leading and at
next-to-leading order. The leading-order result is obtained by
replacing the quantity $K_{\rm NLO}(\delta)$ with
$|C_7^{(0)}(\mu_b)|^2$. The three curves in the upper plot refer to
different choices of $\delta$ (there is no cutoff dependence of the
leading-order result). Note the different scales in the two plots. The
improvement in going from the leading to the next-to-leading order is
spectacular and reduces the apparent scale dependence by more than a
factor of 10. However, a more careful look at the upper plot reveals
two surprises: first, the scale dependence increases rapidly as $\mu_b$
is taken below $0.7 m_b$, although perturbation theory should work well
for lower scales than that; secondly, for $\mu_b>2 m_b$ the prediction
for the partially integrated branching ratio with $\delta=0.3$ exceeds
the prediction for the total branching ratio (obtained with
$\delta=0.9$), which is an unphysical result. Both observations
indicate that higher-order corrections may be more important than what
is suggested by the apparent weak scale dependence of the curves in the
plateau region. The scale dependence of the three most important
contributions to the branching ratio (those from $B_{22}$, $B_{27}$ and
$B_{77}$) is illustrated in the right-hand plots in
Figure~\ref{fig:mub_dep}. There is again a significant improvement in
going from the leading to the next-to-leading order. However, the
residual scale dependence of the quantities $B_{ij}$ at next-to-leading
order is much larger than that of their sum, which determines the total
branching ratio in the Standard Model. Note, in particular, the almost
perfect cancelation of the scale dependence between the 2-2 and the 2-7
terms, which is accidental since the magnitude of the 2-7 term
depends on the top-quark mass through the value of $C_7^{(0)}(m_W)$ in
(\ref{GrSpr}), whereas the 2-2 term is independent of $m_t$. In such a
situation, the apparent weak scale dependence of the sum of all
contributions is not a good measure of higher-order corrections.
Indeed, higher-order corrections must stabilize the different curves in
the right-hand upper plot individually, not only their sum. The
variation of the individual components $B_{ij}$ as a function of
$\mu_b$ thus provides a more conservative estimate of the truncation
error than does the variation of the total branching ratio. For each
component, we estimate the truncation error by taking one half of the
maximum variation obtained by varying $\mu_b$ between $m_b/2$ and
$2m_b$. The truncation error of the sum is then obtained by adding the
individual errors in quadrature. As shown in Table~\ref{tab:errors}, we
find a total truncation error of about $\pm 7\%$, which is more than a
factor of 2 larger than the estimates obtained by previous authors
\cite{Chet}--\cite{Borzu}. In our opinion an even larger truncation
error could be justified given that the choice of the range of
variation of $\mu_b$ is ad hoc, and that the scale dependence of the
various curves in Figure~\ref{fig:mub_dep} is not symmetric around the
point $\mu_b=m_b$.

\begin{table}
\centerline{\parbox{14cm}{\caption{\label{tab:errors}\small\sl
Different sources of theoretical uncertainties (in \%)}}}
\begin{center}
\begin{tabular}{|c|cccccccc|c|}
\hline
$\delta$ & $\mu_b$ & $\bar\mu_b$ & $m_c/m_b$ & $m_b$ & $m_t$
 & $\alpha_s(m_Z)$ & CKM & EW cor.\ & total \\
\hline\hline
0.90 & $\pm 6.3$ & $\mbox{}^{+2.2}_{-1.5}$ & $\mbox{}^{+5.9}_{-5.0}$
 & $\mbox{}^{-1.0}_{+1.1}$ & $\mbox{}^{+1.6}_{-1.7}$
 & $\mbox{}^{+2.8}_{-2.7}$ & $\pm 2.1$ & $\pm 2.0$
 & $\mbox{}^{+10.0}_{-\phantom{1}9.3}$ \\
0.30 & $\pm 6.6$ & $\mbox{}^{+2.2}_{-1.5}$ & $\mbox{}^{+5.5}_{-4.6}$
 & $\mbox{}^{-1.0}_{+1.0}$ & $\mbox{}^{+1.7}_{-1.7}$
 & $\mbox{}^{+2.6}_{-2.6}$ & $\pm 2.1$ & $\pm 2.0$
 & $\mbox{}^{+9.9}_{-9.3}$ \\
0.15 & $\pm 7.1$ & $\mbox{}^{+2.2}_{-1.5}$ & $\mbox{}^{+5.3}_{-4.5}$
 & $\mbox{}^{-0.9}_{+1.0}$ & $\mbox{}^{+1.6}_{-1.7}$
 & $\mbox{}^{+2.5}_{-2.4}$ & $\pm 2.1$ & $\pm 2.0$
 & $\mbox{}^{+10.1}_{-\phantom{1}9.5}$ \\
\hline
\end{tabular}
\end{center}
\end{table}

Let us finally discuss the sensitivity of the theoretical prediction
for the $B\to X_s\gamma$ branching ratio to the various input
parameters entering the calculation. For the quark pole masses, we take
$\sqrt z=m_c/m_b=0.29\pm 0.02$, $m_b=(4.80\pm 0.15)$\,GeV, and
$m_t=(175\pm 6)$\,GeV corresponding to the running mass
$\overline{m}_t(m_W)=(178\pm 6)$\,GeV. We use the two-loop expression
for the running coupling $\alpha_s(\mu)$ with the initial value
$\alpha_s(m_Z)=0.118\pm 0.003$. Finally, for the ratio of the CKM
parameters we take the value $|V_{ts}^* V_{tb}|/|V_{cb}|=0.976\pm
0.010$ obtained from a global analysis of the unitarity triangle
\cite{BurasWar}. We also include an uncertainty of $\pm 2\%$ to account
for next-to-leading electroweak radiative corrections \cite{CzMa}. The
theoretical uncertainties arising from the variation of each input
parameter are collected in Table~\ref{tab:errors}. Adding the different
errors in quadrature we get total uncertainties of about $\pm 10$\% in
all three cases. For the total branching ratio in the Standard Model we
obtain $\mbox{B}(B\to X_s\gamma)=(3.29\pm 0.33)\times 10^{-4} N_{\rm
SL}$. Contrary to common folklore, about 40\% of the branching ratio
reflects the presence of the current--current operator $O_2$ in the
effective Hamiltonian and is not related to penguin diagrams with a
top-quark loop. For comparison with previous authors, we note that with
the choice $\delta=0.99$ we would obtain $\mbox{B}(B\to X_s\gamma)=
(3.37\pm 0.34)\times 10^{-4} N_{\rm SL}$.

\section{Implementation of Fermi motion}
\label{sec:Fermi}

Whereas the explicit power corrections included in the functions
$k_{77}$ and $k_{27}$ in (\ref{kij}) are very small, there is an
important nonperturbative effect that has not been included so far in
next-to-leading order analyses of the $B\to X_s\gamma$ branching ratio:
the residual motion of the $b$ quark inside the $B$ meson caused by its
soft interactions with the light constituents leads to a modification
of the photon energy spectrum, which is an important effect if a
realistic cutoff is imposed \cite{Alex98}. This so-called ``Fermi
motion'' is included in the heavy-quark expansion by resumming an
infinite set of leading-twist corrections into a shape function
$F(k_+)$, which governs the light-cone momentum distribution of the
heavy quark inside the $B$ meson \cite{shape}--\cite{shape2}. This
function shares many similarities with the parton distributions in
deeply inelastic scattering. The physical decay distributions are
obtained from a convolution of parton model spectra with this function.
In the process, phase-space boundaries defined by parton kinematics are
transformed into the proper physical boundaries defined by hadron
kinematics. The shape function is a universal, i.e.\
process-independent characteristic of the $B$ meson governing the
inclusive decay spectra in processes with massless partons in the final
state, such as $B\to X_s\gamma$ and $B\to X_u\,\ell\,\nu$. It is
important to note that this function does not describe in an accurate
way the distributions in decays into massive partons such as $B\to
X_c\,\ell\,\nu$ \cite{Russi,shape2}. Unfortunately, therefore, the
shape function cannot be determined using the lepton spectrum in
semileptonic decays of $B$ mesons, for which high-precision data exist.
On the other hand, there is some useful theoretical information on the
moments of the shape function, which are related to the forward matrix
elements of local operators \cite{shape}:
\begin{equation}
   A_n = \int\mbox{d}k_+\,k_+^n\,F(k_+) = \frac{1}{2 m_B}\,
   \langle B|\,\bar b\,(iD_+)^n b\,|B\rangle \,.
\label{Andef}
\end{equation}
The first three moments satisfy $A_0=1$, $A_1=0$ and $A_2=\frac
13\mu_\pi^2$, where $\mu_\pi^2=-\lambda_1$ is related to the kinetic
energy of the $b$ quark inside the $B$ meson \cite{FaNe}. The condition
$A_1=0$, which is a consequence of the equations of motion, ensures
that the quark mass $m_b$ entering our theoretical expressions is the
pole mass (defined to the appropriate order in perturbation theory,
i.e.\ to one-loop order for our purposes).

Let $P_{\rm p}(y_{\rm p})$ be the photon energy spectrum in the parton
model, where $y_{\rm p}=2 E_\gamma/m_b$ with $0\le y_{\rm p}\le 1$. Our
goal is to include the effects of Fermi motion and calculate the
physical spectrum $P(y)$ as a function of the variable $y=2
E_\gamma/m_B$. To leading-twist approximation, the result is given by
the convolution \cite{shape}
\begin{equation}
   P(y)\,\mbox{d}y = \int\mbox{d}k_+\,F(k_+)\,
   \Big[ P_{\rm p}(y_{\rm p})\,\mbox{d}y_{\rm p}
   \Big]_{y_{\rm p}=y_{\rm p}(k_+)} \,,
\label{convol}
\end{equation}
where $y_{\rm p}(k_+)$ is obtained by replacing $m_b$ in the definition
of $y_{\rm p}$ with the ``effective mass'' $m_b^*=m_b+k_+$
\cite{shape2}, i.e.\ $y_{\rm p}(k_+)=2 E_\gamma/m_b^*=y m_B/m_b^*$.
Because the support of the shape function is restricted to the range
$-m_b\le k_+\le m_B-m_b$, it follows that $0\le y\le 1$. In other
words, after the inclusion of Fermi motion the spectrum extends to the
true kinematic endpoint at $E_\gamma^{\rm max}=m_B/2$. Let us denote by
$B_{\rm p}(\delta_{\rm p})$ the integrated branching ratio calculated
in the parton model, which is given by an integral over the spectrum
$P_{\rm p}(y_{\rm p})$ with a cutoff $\delta_{\rm p}$ defined by the
condition that $E_\gamma\ge\frac 12(1-\delta_{\rm p}) m_b$. From
(\ref{convol}), it then follows that the corresponding physical
quantity $B(\delta)$ with $\delta$ defined such that $E_\gamma\ge\frac
12(1-\delta) m_B$ is given by
\begin{equation}
   B(\delta) = \int\limits_{m_B(1-\delta)-m_b}^{m_B-m_b}\!
   \mbox{d}k_+\,F(k_+)\,B_{\rm p}\!\left(
   1 - \frac{m_B(1-\delta)}{m_b+k_+} \right) \,.
\label{recipe}
\end{equation}
This relation is such that $B(1)=B_{\rm p}(1)$, implying that the total
branching ratio is not affected by Fermi motion; indeed, the $1/m_Q^2$
corrections in (\ref{kij}) are the only power corrections to the total
branching ratio. The effects of Fermi motion are, however, important
for realistic values of the cutoff. We will now evaluate relation
(\ref{recipe}) for the various components $B_{ij}(\delta)$ introduced
in the previous section.

\begin{table}
\centerline{\parbox{14cm}{\caption{\label{tab:kij_Fermi}\small\sl
Values of the coefficients $B_{ij}(\delta)$ in units of $10^{-4}$,
corrected for Fermi motion}}}
\begin{center}
\begin{tabular}{|ccc|cccccc|c|}
\hline
$m_b$~[GeV] & $\delta$ & $E_\gamma^{\rm min}~[{\rm GeV}]$ & $B_{22}$
 & $B_{77}$ & $B_{88}$ & $B_{27}$ & $B_{28}$ & $B_{78}$
 & $\sum B_{ij}$ \\
\hline\hline
4.65 & 0.90 & 0.26 & 1.289 & 0.375 & 0.014 & 1.401 & 0.162 & 0.083
 & 3.324 \\
 & 0.30 & 1.85 & 1.201 & 0.333 & 0.004 & 1.322 & 0.132 & 0.075
 & 3.068 \\
 & 0.15 & 2.24 & 0.802 & 0.220 & 0.002 & 0.889 & 0.087 & 0.050
 & 2.050 \\
\hline
4.80 & 0.90 & 0.26 & 1.258 & 0.382 & 0.014 & 1.395 & 0.160 & 0.083
 & 3.291 \\
 & 0.30 & 1.85 & 1.213 & 0.352 & 0.004 & 1.362 & 0.134 & 0.078
 & 3.144 \\
 & 0.15 & 2.24 & 0.945 & 0.272 & 0.003 & 1.071 & 0.104 & 0.060
 & 2.456 \\
\hline
4.95 & 0.90 & 0.26 & 1.227 & 0.388 & 0.014 & 1.388 & 0.157 & 0.083
 & 3.259 \\
 & 0.30 & 1.85 & 1.200 & 0.365 & 0.004 & 1.375 & 0.133 & 0.080
 & 3.156 \\
 & 0.15 & 2.24 & 1.072 & 0.323 & 0.003 & 1.239 & 0.118 & 0.070
 & 2.825 \\
\hline
\end{tabular}
\end{center}
\end{table}

Several ans\"atze for the shape function have been suggested in the
literature \cite{shape}--\cite{shape2}. For our purposes, given the
poor present knowledge about higher moments $A_n$ with $n\ge 3$, it is
sufficient to adopt the simple form
\begin{equation}
   F(k_+) = N\,(1-x)^a e^{(1+a)x} \,;\quad
   x = \frac{k_+}{\bar\Lambda} \le 1 \,,
\label{shapex}
\end{equation}
where $\bar\Lambda=m_B-m_b$. This ansatz is such that $A_1=0$ by
construction,\footnote{For simplicity, we neglect exponentially small
terms in $m_b/\bar\Lambda$.}
whereas the condition $A_0=1$ fixes the normalization $N$. The
parameter $a$ can be related to the second moment, yielding $A_2=\frac
13\mu_\pi^2=\bar\Lambda^2/(1+a)$. Thus, the $b$-quark mass (or
$\bar\Lambda$) and the quantity $\mu_\pi^2$ (or $a$) are the two
parameters of our function. Below, we will take $m_b=4.8$\,GeV and
$\mu_\pi^2=0.3$\,GeV$^2$ as reference values, in which case $a\approx
1.29$. Because the main effect of Fermi motion is to fill the gap
between the parton model endpoint of the photon spectrum and the
physical endpoint, it turns out that the results are very sensitive to
the choice of the $b$-quark mass. Table~\ref{tab:kij_Fermi} shows the
coefficients $B_{ij}(\delta)$ corrected for Fermi motion using the
above ansatz with a fixed value $a\approx 1.29$ but different values of
$m_b$. A representative range of parameters is covered by considering
the three following cases: $m_b=4.65$\,GeV (yielding
$\bar\Lambda\approx 0.63$\,GeV and $\mu_\pi^2\approx 0.52$\,GeV$^2$),
$m_b=4.8$\,GeV (yielding $\bar\Lambda\approx 0.48$\,GeV and
$\mu_\pi^2\approx 0.3$\,GeV$^2$), and $m_b=4.95$\,GeV (yielding
$\bar\Lambda\approx 0.33$\,GeV and $\mu_\pi^2\approx 0.14$\,GeV$^2$).
We also show the values of the photon energy cutoff, $E_\gamma^{\rm
min}=\frac 12(1-\delta) m_B$, which are now independent of the
$b$-quark mass.

\begin{figure}
\epsfxsize=14cm
\centerline{\epsffile{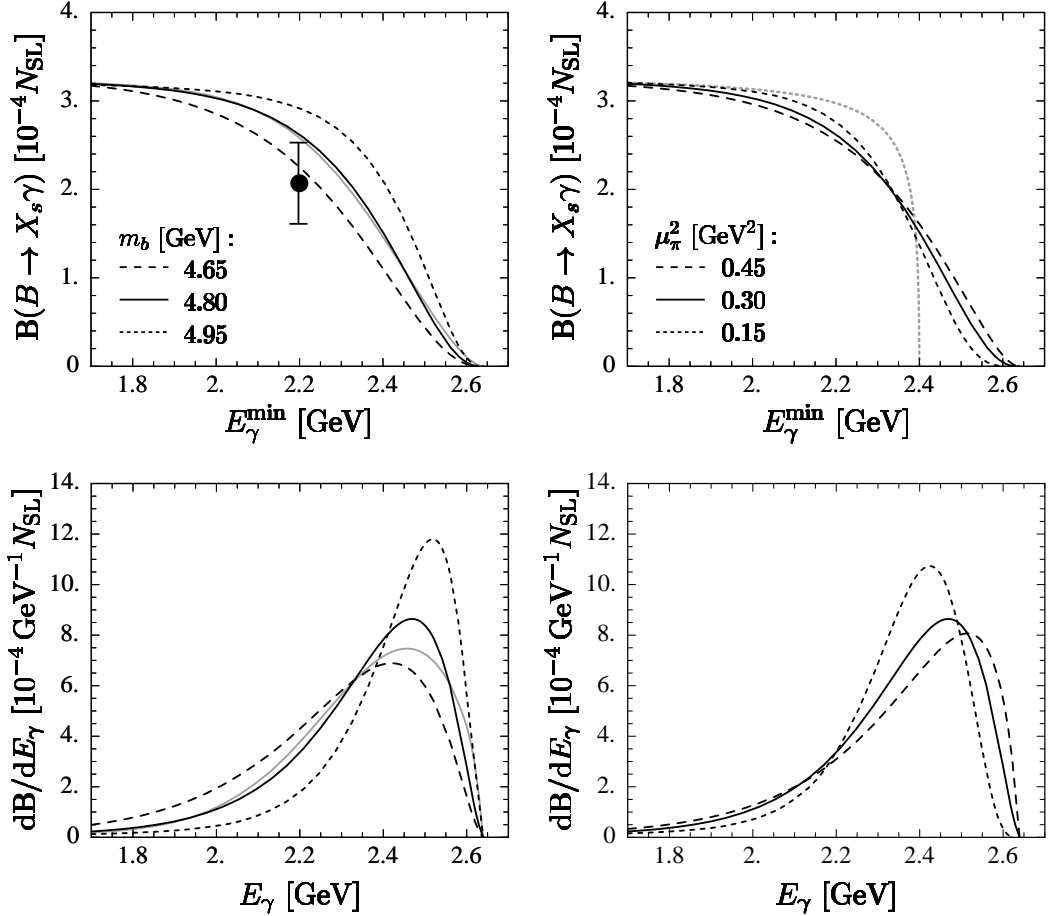}}
\centerline{\parbox{14.5cm}{\caption{\label{fig:Fermi_motion}\small\sl
Theoretical predictions for the integrated $B\to X_s\gamma$ branching
ratio (upper plots) and the corresponding photon spectra (lower plots)
for various choices of the shape-function parameters ($m_b$,
$\mu_\pi^2$) and functional form, as explained in the text. The
calculation of the photon spectra will be discussed in
Section~\protect\ref{sec:spectrum}.}}}
\end{figure}

For a graphical illustration of the sensitivity of our results to the
parameters of the shape function, we show in the upper plots in
Figure~\ref{fig:Fermi_motion} the predictions for the Standard Model
branching ratio as a function of the energy cutoff $E_\gamma^{\rm
min}$. In the first plot, we use the same sets of parameters as in
Table~\ref{tab:kij_Fermi}, i.e.\ $m_b=4.65$\,GeV (long-dashed curve),
4.8\,GeV (solid curve), and 4.95\,GeV (short-dashed curve), with
$\mu_\pi^2$ adjusted such that the ratio $\mu_\pi^2/\bar\Lambda^2$
remains constant. The gray line shows the result obtained using the
same parameters as for the solid line, but with a Gaussian ansatz
$F(k_+)=N\,(1-x)^a e^{-b(1-x)^2}$ for the shape function. For
comparison, we also show the data point $\mbox{B}(B\to
X_s\gamma)=(2.04\pm 0.47)\times 10^{-4}$ obtained by the CLEO
Collaboration with a cutoff at 2.2\,GeV \cite{private}. As shown in
Appendix~B, the fact that in the CLEO analysis the cutoff is imposed on
the photon energy in the laboratory frame rather than in the rest frame
of the $B$ meson is not very important for the partially integrated
branching ratio and has been neglected here. In the second plot, we
keep $m_b=4.8$\,GeV fixed and compare the parton model result (gray
dotted curve) with the results corrected for Fermi motion, using
$\mu_\pi^2=0.15$\,GeV$^2$ (short-dashed curve), 0.30\,GeV$^2$ (solid
curve), and 0.45\,GeV$^2$ (long-dashed curve). This figure illustrates
how Fermi motion fills the gap between the parton model endpoint at
$m_b/2$ and the physical endpoint at $m_B/2$. To be precise, the
physical endpoint is actually located at $[m_B^2-(m_K+m_\pi)^2]/2
m_B\approx 2.60$\,GeV, i.e.\ slightly below $m_B/2\approx 2.64$\,GeV.
Close to the endpoint, our theoretical prediction provides an average
description of the true spectrum in the sense of quark--hadron duality
(see Section~\protect\ref{sec:mass}). Comparing the two upper plots in
Figure~\ref{fig:Fermi_motion}, we observe that the uncertainty due to
the value of the $b$-quark mass is the dominant one. Variations of the
parameter $\mu_\pi^2$ have a much smaller effect on the partially
integrated branching ratio, and also the sensitivity to the functional
form adopted for the shape function turns out to be small. This
behaviour is a consequence of global quark--hadron duality, which
ensures that even partially integrated quantities are rather
insensitive to bound-state effects. The strong remaining dependence on
the $b$-quark mass is simply due to the transformation by Fermi motion
of phase-space boundaries from parton to hadron kinematics. We believe
that the spread of results obtained by varying $m_b$ between 4.65 and
4.95\,GeV (with $\mu_\pi^2$ adjusted as described above) is a fair
representation of the amount of model dependence resulting from the
inclusion of Fermi motion. With a cutoff $E_\gamma^{\rm min}=2.2$\,GeV
as used in the CLEO analysis, and correcting for the small effect of
the boost from the $B$ rest frame to the laboratory frame (see
Appendix~B), we obtain
\begin{equation}
   {\rm B}(B\to X_s\gamma)\big|_{E_\gamma^{\rm lab}>2.2\,{\rm GeV}}
   = (2.57\pm 0.26_{-0.36}^{+0.31})\times 10^{-4} N_{\rm SL} \,.
\end{equation}
The first error accounts for the dependence on the various input
parameters, while the second one reflects the uncertainty due to the
modeling of Fermi motion. For a cutoff as high as that employed in the
CLEO analysis, this uncertainty is in fact the dominant theoretical
error. In the future, an effort should therefore be made to lower the
cutoff on the photon energy to a value of 2\,GeV or less. Comparing our
result with the CLEO measurement of $(2.04\pm 0.47)\times 10^{-4}$
\cite{private}, we obtain the ratio
$R={\rm B_{exp}/B_{th}}=0.79\pm 0.18({\rm exp})\mbox{}^{+0.14}_{-0.12}
({\rm th})$ (assuming $N_{\rm SL}=1$), which deviates from unity by
less than one standard deviation. This must be confronted with the
comparison of the total branching ratios using the value reported in
(\ref{CLEObr}), which gives $R={\rm B_{exp}/B_{th}}=0.71\pm 0.20({\rm
exp})\pm 0.07({\rm th})$. The extrapolation to low photon energies
performed in the CLEO analysis \cite{CLEO} has artificially increased
the deviation from the theoretical prediction by a significant amount,
i.e.\ the model dependence inherent in this extrapolation has been
underestimated. This is also reflected in the fact that the correction
factor relating the total branching ratio to the branching ratio
obtained with the cut $E_\gamma^{\rm lab}>2.2$\,GeV is
$K_{2.2}=0.78_{-0.11}^{+0.09}$, which differs significantly from the
factor $0.87\pm 0.06$ employed by CLEO. Using our correction factor,
the extrapolation of the CLEO measurement to the total branching ratio
yields
\begin{equation}
   {\rm B}(B\to X_s\gamma)_{\rm CLEO} = (2.62\pm 0.60_{\rm exp}\,
   \mbox{}^{+0.37}_{-0.30\,{\rm th}})\times 10^{-4}
\label{newCLEO1}
\end{equation}
instead of the value quoted in (\ref{CLEObr}). We stress that the
change in the central value and the increase of the theoretical
error\footnote{The second error quoted in (\protect\ref{CLEObr}) is
dominated by experimental systematics. The theoretical error was
assumed to be $\pm 0.16\times 10^{-4}$.}
with respect to the result reported by CLEO are entirely due to the
improved treatment of bound-state effects presented in this paper.
Whereas the CLEO analysis relies on a quark model, we perform an
analysis that is entirely based on QCD and the operator product
expansion. Our treatment is thus not only more conservative but also
more consistent from a theoretical point of view.

The procedure of extrapolating a measurement of the $B\to X_s\gamma$
branching ratio in the region of high photon energies to the total
branching ratio not only introduces large systematic errors, but also
entails the disadvantage that one has to rely on the Standard Model to
describe the photon spectrum in the low-energy region. In our opinion,
it would therefore be desirable if in the future the comparison of
theory with experiment were done for the partially integrated branching
ratio, which is the quantity actually measured, rather than for the
total branching ratio.

\boldmath
\section{Photon spectrum and determination of $m_b$}
\label{sec:spectrum}
\unboldmath

The large theoretical uncertainties in the calculation of Fermi motion
effects on the partially integrated branching ratio for $B\to
X_s\gamma$ decays can be reduced in two ways. The first possibility is
to lower the cutoff $E_\gamma^{\rm min}$ on the photon energy. As is
apparent from Figure~\ref{fig:Fermi_motion}, if a value as low as
$E_\gamma^{\rm min}\lsim 2$\,GeV could be achieved, the theoretical
predictions would become insensitive to the parameters of the shape
function. To what extent this will be possible in future experiments
depends on their capability to reject the background of photons from
other decays. The Cabibbo-favoured $B$ decays into charmed particles,
in particular, can yield photons of energy up to about 2.3\,GeV. The
second possibility is that future high-precision measurements of the
photon spectrum will make it possible to adjust the parameters of the
shape function from a fit to the data. We repeat that these parameters
cannot be determined from a study of the lepton spectrum in $B\to
X_c\,\ell\,\nu$ decays. On the other hand, a determination of the
shape-function parameters from $B\to X_s\gamma$ decays would enable us
to predict the lepton spectrum in $B\to X_u\,\ell\,\nu$ in a
model-independent way \cite{shape}. This may help to reduce the
theoretical uncertainty in the current value of $|V_{ub}|$. A detailed
analysis of the photon spectrum will therefore be an important aspect
in future analyses of inclusive radiative $B$ decays.

Given the expression for the integrated $B\to X_s\gamma$ branching
ratio derived in the previous sections, the photon spectrum can be
obtained from differentiation with respect to $\delta$, i.e.
\begin{equation}
   P(y) = \frac{1}{\mbox{B}(B\to X_c\,e\,\bar\nu)}\,
   \frac{\mbox{dB}(B\to X_s\gamma)}{\mbox{d}y}
   = \frac{6\alpha}{\pi f(z)}\,\left| \frac{V_{ts}^* V_{tb}}{V_{cb}}
    \right|^2\,K_{\rm NLO}'(1-y) \,,
\end{equation}
where $y=E_\gamma/E_\gamma^{\rm max}$. In analogy with (\ref{KNLO}), we
write
\begin{eqnarray}
   &&\hspace{-0.5cm}
   K_{\rm NLO}'(1-y) \nonumber\\
   &=& \sum_{ \stackrel{i,j=2,7,8}{i\le j} }
    p_{ij}(y,\mu_b)\,\mbox{Re}\!\left[ C_i^{(0)}(\mu_b)\,
    C_j^{(0)*}(\mu_b) \right] +
\Delta(y)\,\frac{\alpha_s(\mu_b)}{2\pi}\,
    \mbox{Re}\!\left[ C_7^{(1)}(\mu_b)\,C_7^{(0)*}(\mu_b) \right]
    \nonumber\\
   &&\mbox{}+ \Delta(y)\,\frac{\alpha}{\alpha_s(\mu_b)} \bigg(
    2\,\mbox{Re}\!\left[ C_7^{({\rm em})}(\mu_b)\,C_7^{(0)*}(\mu_b)
    \right] - k_{\rm SL}^{({\rm em})}(\mu_b)\,|C_7^{(0)}(\mu_b)|^2
    \bigg) \,,
\label{PNLO}
\end{eqnarray}
where
\begin{equation}
   \Delta(y) = -\frac{4\alpha_s(\mu_b)}{3\pi(1-y)}
   \left( \ln(1-y) + \frac 74 \right)
   \exp\!\left[ -\frac{2\alpha_s(\mu_b)}{3\pi} \left( \ln^2(1-y)
   + \frac 72\ln(1-y) \right) \right]
\label{Deltadef}
\end{equation}
is the derivative of the Sudakov factor. The Sudakov factor can be
regarded as a smeared step function, and hence $\Delta(y)$ can be
viewed as a smeared $\delta$-function. In other words, we must consider
$\Delta(y)=O(1)$ rather than $O(\alpha_s)$, in spite of the prefactor
$\alpha_s$ in (\ref{Deltadef}). The coefficient functions
$p_{ij}(y,\mu_b)$ in (\ref{PNLO}) have the same form as the
coefficients $k_{ij}(\delta,\mu_b)$ in (\ref{kij}) but with the
replacements $S(\delta)\to\Delta(y)$ and $f_{ij}(\delta)\to s_{ij}(y)$,
where $s_{ij}(y)=f_{ij}'(1-y)$. The explicit expressions for these
functions are presented in Appendix~C.

\begin{figure}
\epsfxsize=14cm
\centerline{\epsffile{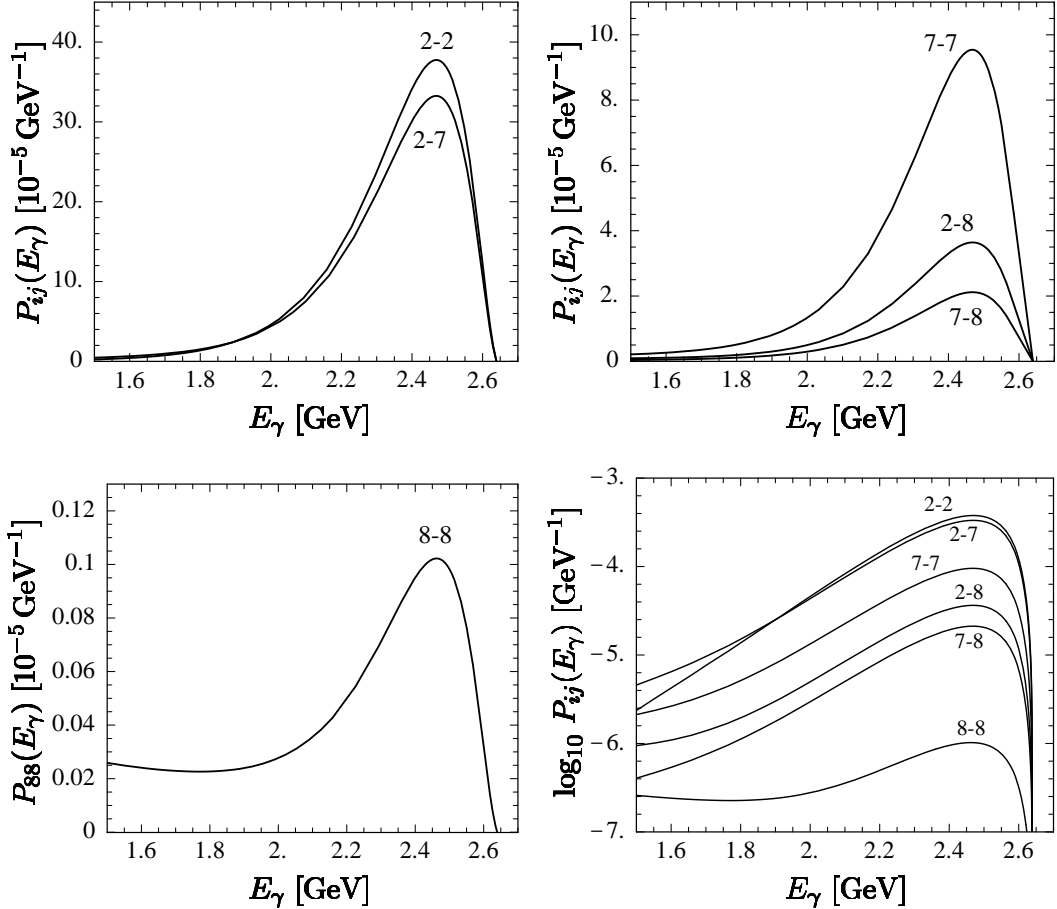}}
\centerline{\parbox{14.5cm}{\caption{\label{fig:spectra}\small\sl
Different components of the photon spectrum in $B\to X_s\gamma$
decays}}}
\end{figure}

Given the photon spectrum $P_{\rm p}(y_{\rm p})$ in the parton model,
where $y_{\rm p}=2 E_\gamma/m_b$, the next step is to implement Fermi
motion. According to (\ref{convol}), this is achieved by taking the
convolution
\begin{equation}
   P(y) = \int\limits_{m_B y-m_b}^{m_B-m_b}\!\mbox{d}k_+\,
   F(k_+)\,\frac{m_B}{m_b+k_+}\,P_{\rm p}\!\left(
   \frac{m_B y}{m_b+k_+} \right) \,,
\label{py}
\end{equation}
where $y=2 E_\gamma/m_B$. It is convenient to decompose the final
result for the spectrum in a form analogous to (\ref{xidecomp}) by
writing
\begin{eqnarray}
   &&\hspace{-0.5cm}
    \frac{1}{N_{\rm SL}}\,
    \frac{\mbox{dB}(B\to X_s\gamma)}{\mbox{d}E_\gamma}
    = 0.105\times\frac{2}{m_B}\,P(2 E_\gamma/m_B)  \nonumber\\
   &=& P_{22}(E_\gamma) + P_{77}(E_\gamma)
    \left( |\xi_7|^2 + |\xi_7^R|^2 \right) + P_{88}(E_\gamma)
    \left( |\xi_8|^2 + |\xi_8^R|^2 \right) \nonumber\\
   &&\mbox{}+ P_{27}(E_\gamma)\,\mbox{Re}(\xi_7)
    + P_{28}(E_\gamma)\,\mbox{Re}(\xi_8)
    + P_{78}(E_\gamma) \left[ \mbox{Re}(\xi_7\xi_8^*)
    + \mbox{Re}(\xi_7^R\xi_8^{R*}) \right] \,.
\label{Sdecomp}
\end{eqnarray}
The results for the various components of the spectrum are shown in
Figure~\ref{fig:spectra}, where we take central values of all input
parameters. The contributions are ordered according to their magnitude.
In the last plot, we show all components together on a logarithmic
scale. Note that, with the exception of the tiny 8-8 contribution, the
different components have a very similar spectral shape. This
observation implies that the shape of the photon spectrum is not
sensitive to physics beyond the Standard Model. With a realistic cutoff
on the photon energy, even large deviations of the parameters
$\xi_7^{(R)}$ and $\xi_8^{(R)}$ from their standard values would not
have a detectable effect on the shape of the photon spectrum. Although
this may be disappointing from the point of view of searching for New
Physics in $B\to X_s\gamma$ decays, it entails the advantage that a
precise measurement of the spectrum can be used to determine the
parameters of the shape function without relying on the Standard Model.
For the remainder of this section we concentrate on the Standard Model,
for which the photon spectrum is given by the sum of the individual
contributions shown in Figure~\ref{fig:spectra}. The results obtained
for various choices of the parameters of the shape function are shown
in the lower plots in Figure~\ref{fig:Fermi_motion}. The photon spectra
are more sensitive to the functional form of the shape function than
are the predictions for the integrated branching ratio in the upper
plots. Therefore, a fit to future high-precision data on the spectrum
should use a more flexible ansatz for the shape function than the one
given in (\ref{shapex}). On the other hand, we will see below that even
a small element of smearing provided, e.g., by the finite detector
resolution or the Lorentz boost of photons from the $B$ rest frame to
the laboratory frame, is sufficient to reduce this sensitivity
significantly.

\begin{figure}
\epsfxsize=14cm
\centerline{\epsffile{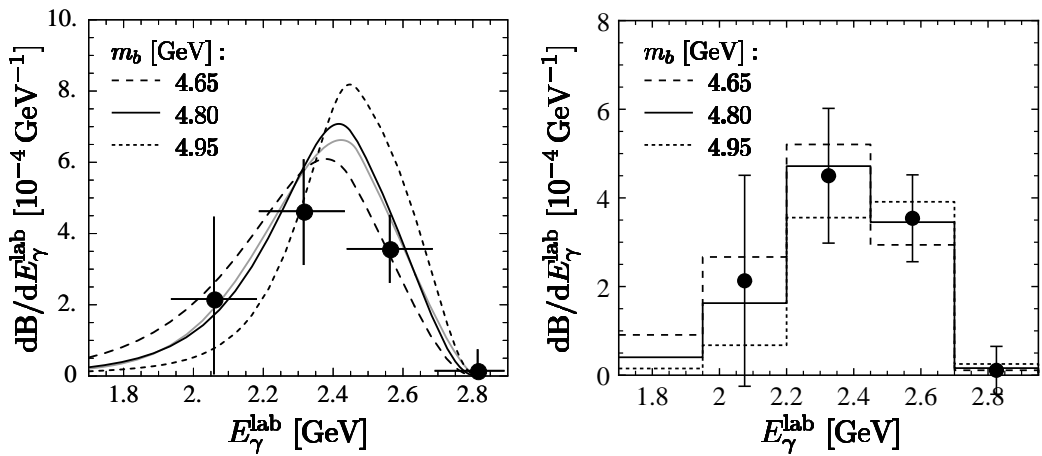}}
\centerline{\parbox{14.5cm}{\caption{\label{fig:Elab}\small\sl
Theoretical predictions for the photon energy spectrum in the
laboratory frame for different parameters of the shape function. The
gray line in the left-hand plot shows the result obtained using a
Gaussian form of the shape function with central values of $m_b$ and
$\mu_\pi^2$. The data points show the CLEO results. In the left-hand
plot, no fit to the data is performed, whereas the right-hand plot
shows the results of the best fits reported in
Table~\protect\ref{tab:fits}.}}}
\end{figure}

Let us now make a comparison of our predictions for the photon energy
spectrum with the data obtained by the CLEO Collaboration, which are
presented in Table~\ref{tab:CLEOdata}. To this end, we must take into
account that $B$ mesons produced in decays of the $\Upsilon(4s)$
resonance have a small momentum in the laboratory system, so that the
photon spectrum is Doppler shifted. The boost that connects the $B$
rest frame with the laboratory frame is characterized by
\begin{equation}
   \beta = \frac{|{\bf p}_B|}{E_B}
   = \sqrt{1-\frac{4 m_B^2}{m_{\Upsilon(4s)}^2}}
   = 0.064\pm 0.007 \,,
\end{equation}
and the maximum energy in the laboratory frame is $(E_\gamma^{\rm
lab})_{\rm max}\approx (1+\beta) E_\gamma^{\rm max}\approx
E_\gamma^{\rm max} + 170$\,MeV. The formalism for incorporating this
effect is discussed in Appendix~B. The results are shown in the
left-hand plot in Figure~\ref{fig:Elab}, where we compare the corrected
photon spectra with the CLEO data, using our standard parameters for
the shape function. No fit to the data has been performed; in all
cases, the theoretical spectra are normalized to the central Standard
Model value of $3.29\times 10^{-4}$ for the total branching ratio
(assuming $N_{\rm SL}=1$). Note that after the smearing implied by the
Doppler shift of the spectra, the gray and the black solid lines, which
as before correspond to different functional forms adopted for the
shape function, are very close together. This reflects the reduced
sensitivity to the fine details of the modeling of Fermi motion, which
is achieved by any kind of smearing of the photon spectrum.

\begin{table}
\centerline{\parbox{14cm}{\caption{\label{tab:CLEOdata}\small\sl CLEO
results for the photon spectrum in $B\to X_s\gamma$ decays
\protect\cite{private}. The value of the branching ratio reported in
\protect\cite{CLEO} was obtained from the middle two bins.}}}
\begin{center}
\begin{tabular}{|cc|}
\hline
$\Delta E_\gamma^{\rm lab}~[{\rm GeV}]$
 & ${\rm dB}/{\rm d}E_\gamma^{\rm lab}~[10^{-4}\,{\rm GeV}^{-1}]$ \\
\hline\hline
1.95--2.20 & $2.13\pm 2.38$ \\
2.20--2.45 & $4.50\pm 1.52$ \\
2.45--2.70 & $3.54\pm 0.98$ \\
2.70--2.95 & $0.11\pm 0.54$ \\
\hline
\end{tabular}
\end{center}
\end{table}

To perform a fit to the data, we rebin our theoretical results in the
same energy intervals as used by CLEO and, for each set of parameters
for the shape function, adjust the overall normalization (i.e.\ the
total branching ratio) to give the best fit to the data. The results
are reported in Table~\ref{tab:fits}, and the best fits displayed in
the right-hand plot in Figure~\ref{fig:Elab}. All fits have an
excellent $\chi^2/n_{\rm dof}\ll 1$, indicating that with the present
accuracy of the data it is not possible to determine the parameters of
the shape function in a meaningful way. Unless a very large value of
$m_b$ is chosen, the result for the total branching ratio comes out
higher than the value (\ref{CLEObr}) reported by CLEO, and the upper
bounds for the branching ratio obtained at 90\% confidence level are
well above the Standard Model prediction of $(3.29\pm 0.33)\times
10^{-4}$. Combining the results obtained for the three choices of
$m_b$, we get
\begin{equation}
   {\rm B}(B\to X_s\gamma)_{\rm CLEO}^{\rm spectrum}
   = (2.66\pm 0.56_{\rm exp}\,\mbox{}^{+0.43}_{-0.48\,{\rm th}})
   \times 10^{-4} \,,
\label{newCLEO2}
\end{equation}
corresponding to a ratio $R={\rm B_{exp}/B_{th}}=0.81\pm 0.17({\rm
exp})\mbox{}^{+0.13}_{-0.15}({\rm th})$, which deviates from unity by
less than one standard deviation. In quoting the theoretical error we
assume that the spread of the results shown in the three columns of
Table~\ref{tab:fits} represents a reasonable estimate of the
theoretical uncertainty arising from the modeling of the shape
function. The individual $R$ ratios for the different choices of $m_b$
are displayed in the last column in the table. The result
(\ref{newCLEO2}) is in good agreement with the value (\ref{newCLEO1})
obtained from the extrapolation of the partially integrated branching
ratio measured by CLEO with a photon-energy cutoff at 2.2\,GeV. Since
the fit to the photon spectrum uses more experimental information, we
tend to consider (\ref{newCLEO2}) to be the more conservative result of
the two. We will therefore use this value in our further analysis.

\begin{table}
\centerline{\parbox{14cm}{\caption{\label{tab:fits}\small\sl
Results for the total $B\to X_s\gamma$ branching ratio, and upper
limits at 90\% confidence level, obtained from a fit to the CLEO data
shown in Table~\protect\ref{tab:CLEOdata}. The quantity $R$ denotes the
ratio of the extracted value for the branching ratio to the Standard
Model prediction.}}}
\begin{center}
\begin{tabular}{|c|ccc|}
\hline
 & $m_b=4.65$\,GeV & $m_b=4.80$\,GeV & $m_b=4.95$\,GeV \\
\hline\hline
${\rm B}(B\to X_s\gamma)~[10^{-4}]$ & $3.09\pm 0.66$ & $2.66\pm 0.56$
 & $2.18\pm 0.47$ \\
90\% CL & $<4.66$ & $<4.06$ & $<3.26$ \\
$\chi^2/n_{\rm dof}$ & 0.64/3 & 0.08/3 & 0.97/3 \\
$R$ & $0.94\pm 0.20\pm 0.09$ & $0.81\pm 0.17\pm 0.08$
 & $0.66\pm 0.14\pm 0.07$ \\
\hline
\end{tabular}
\end{center}
\end{table}

\begin{figure}
\epsfxsize=14cm
\centerline{\epsffile{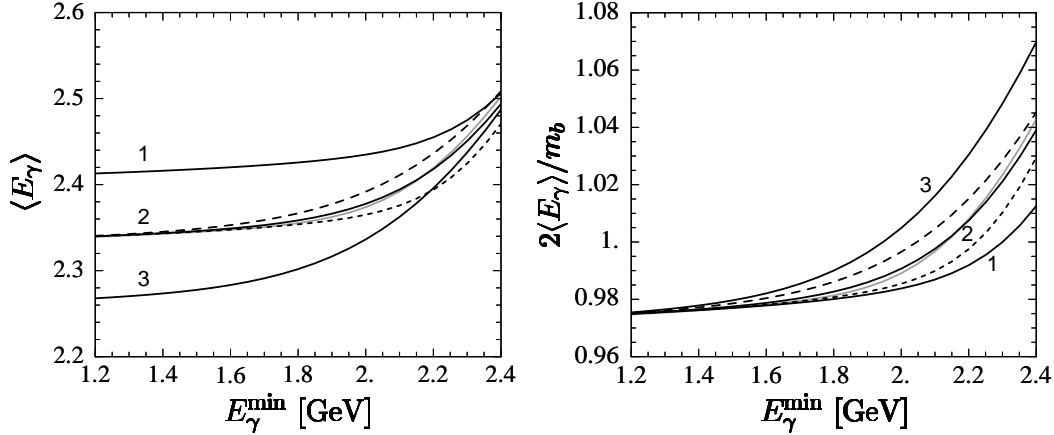}}
\centerline{\parbox{14.5cm}{\caption{\label{fig:mb}\small\sl
Theoretical predictions for the average photon energy as a function of
the cutoff $E_\gamma^{\rm min}$, for different parameters of the shape
function. The solid lines refer to: (1) $m_b=4.95$\,GeV and
$\mu_\pi^2=0.14$\,GeV$^2$; (2) $m_b=4.8$\,GeV and
$\mu_\pi^2=0.3$\,GeV$^2$; (3) $m_b=4.65$\,GeV and
$\mu_\pi^2=0.52$\,GeV$^2$. The dashed lines show the dependence on the
value of $\mu_\pi^2$ for the case $m_b=4.8$\,GeV, where
$\mu_\pi^2=0.3$\,GeV$^2$ (solid), 0.45\,GeV$^2$ (long-dashed),
0.15\,GeV$^2$ (short-dashed). The gray lines refer to the Gaussian
ansatz for the shape function.}}}
\end{figure}

Once high-statistic measurements of the photon spectrum become
available, it will be possible to determine the parameters of the shape
function directly from the data. In particular, the average photon
energy in $B\to X_s\gamma$ decays is a sensitive measure of the
$b$-quark mass. In practice, what can be measured is the average photon
energy as a function of the cutoff $E_\gamma^{\rm min}$, given by
\begin{equation}
   \langle E_\gamma\rangle
   = \frac{\int\limits_{E_\gamma^{\rm min}}^{E_\gamma^{\rm max}}\!
      {\rm d}E_\gamma\,E_\gamma\,
      \displaystyle\frac{\rm dB}{{\rm d}E_\gamma}}
     {\int\limits_{E_\gamma^{\rm min}}^{E_\gamma^{\rm max}}\!
      {\rm d}E_\gamma\,\displaystyle\frac{\rm dB}{{\rm d}E_\gamma}} \,.
\label{Eavg}
\end{equation}
Provided that $E_\gamma^{\rm min}$ is not too close to the endpoint,
this quantity is insensitive to the details of the shape function
except for the value of $m_b$. Indeed, at leading order in the
heavy-quark expansion one simply gets $\langle E_\gamma\rangle=m_b/2$.
The result for the average photon energy obtained by including the full
next-to-leading order corrections is shown in Figure~\ref{fig:mb}. For
simplicity, we neglect in this figure the boost between the $B$ rest
frame and the laboratory frame, which has a very small effect on the
average photon energy. The different curves in each plot refer to the
various sets of shape-function parameters considered previously in
Figure~\ref{fig:Fermi_motion}. In the right-hand plot we show the
average photon energy normalized to $m_b/2$. We observe that for
$E_\gamma^{\rm min}\lsim 1.8$\,GeV the mean photon energy provides a
sensitive measure of the mass of the $b$ quark, which to a good
approximation is independent of other shape-function parameters such as
$\mu_\pi^2$. Asymptotically, for very small cutoff values $\langle
E_\gamma\rangle$ is lower than $m_b/2$ by about 3\%. For $E_\gamma^{\rm
min}\gsim 1.8$\,GeV, on the other hand, the sensitivity to the
modeling of Fermi motion quickly increases.  
For a cutoff at 2.2\,GeV
as employed in the CLEO analysis, there is very little sensitivity to
the value of $m_b$.  

Moments of the type shown in (\ref{Eavg}) have been considered
previously by Kapustin and Ligeti \cite{Kapu} employing the heavy-quark
expansion but not including the effects of Fermi motion. These authors
make predictions for the average photon energy and for the width of the
photon spectrum for cutoff values between 1.8 and 2.2\,GeV. Our
findings show that for $E_\gamma^{\rm min}>1.8$\,GeV there are
significant corrections to these predictions caused by Fermi
motion.  This has also been noted by Bauer \cite{Bauer}.

Let us assume that in the future it will be possible to measure the
average photon energy in $B\to X_s\gamma$ decays with a cutoff low
enough to be insensitive to Fermi motion effects. It is instructive to
understand the precise meaning of the $b$-quark mass that can then be
extract from the data. In the limit where Fermi motion can be
neglected, it follows from (\ref{py}) that
\begin{equation}
   \langle E_\gamma\rangle = \frac{m_b}{2}\times
   \frac{\int\limits_{1-\delta_{\rm p}}^1\!{\rm d}y_{\rm p}\,
         y_{\rm p}\,P_{\rm p}(y_{\rm p})}
        {\int\limits_{1-\delta_{\rm p}}^1\!{\rm d}y_{\rm p}\,
         P_{\rm p}(y_{\rm p})} \,,
\end{equation}
where $m_b$ is the pole mass, $P_{\rm p}(y_{\rm p})$ is the photon
spectrum in the parton model, and $1-\delta_{\rm p}=2 E_\gamma^{\rm
min}/m_b$. Using that
\begin{equation}
   \int\limits_{1-\delta_{\rm p}}^1\!{\rm d}y_{\rm p}\,y_{\rm p}\,
   P_{\rm p}(y_{\rm p})
   = \int\limits_0^{\delta_{\rm p}}\!{\rm d}z
   \int\limits_{1-z}^1\!{\rm d}y_{\rm p}\,P_{\rm p}(y_{\rm p})
   + (1-\delta_{\rm p})
   \int\limits_{1-\delta_{\rm p}}^1\!{\rm d}y_{\rm p}\,
   P_{\rm p}(y_{\rm p}) \,,
\end{equation}
we find for the average photon energy
\begin{equation}
   \langle E_\gamma\rangle = \frac{m_b}{2}\,C_E[\alpha_s(m_b)]
   \left\{ 1 + \frac{\alpha_s(m_b)}{\pi}\,D(\delta_{\rm p})
   + \delta_{\rm HT} + \dots \right\} \,,
\label{Egamavg}
\end{equation}
where
\begin{eqnarray}
   C_E[\alpha_s] &=& 1 -\frac{23}{54}\,\frac{\alpha_s}{\pi}
    + O(\alpha_s^2) \approx 0.97 \,, \nonumber\\
   D(\delta_{\rm p}) &=& \bar d_{77}(\delta_{\rm p})
    + \sum_{ \stackrel{i,j=2,7,8}{i\le j} }\mbox{}^{\!\!\!\!\!\!\prime}
    ~\,d_{ij}(\delta_{\rm p})\,
    \frac{{\rm Re}[C_i^{(0)}(m_b) C_j^{(0)*}(m_b)]}
         {|C_7^{(0)}(m_b)|^2} \,.
\end{eqnarray}
The prime on the sum indicates that $(i,j)\ne(7,7)$. The functions
$d_{ij}(\delta_{\rm p})$ and $\bar d_{77}(\delta_{\rm p})$ are
collected in Appendix~C. The normalization of $\bar d_{77}$ is such
that $\bar d_{77}(1)=0$, i.e.\ apart from corrections due to operator
mixing the average photon energy without any cutoff is given by
$\frac12 m_b C_E[\alpha_s]$. Finally, the quantity $\delta_{\rm HT}$ in
(\ref{Egamavg}) parametrizes ``higher-twist corrections'', which are
neglected in our leading-twist approximation to the shape function. An
explicit calculation of these corrections gives \cite{Adam,shape}
\begin{equation}
   \delta_{\rm HT} = - \frac{\lambda_1+3\lambda_2}{2 m_b^2}
   \approx (-0.1\pm 0.4)\% \,,
\end{equation}
where we have assumed that $-\lambda_1=\mu_\pi^2=(0.3\pm
0.2)$\,GeV$^2$. The contribution of $\delta_{\rm HT}$ in
(\ref{Egamavg}) has a negligible effect.

\begin{figure}
\epsfxsize=8cm
\centerline{\epsffile{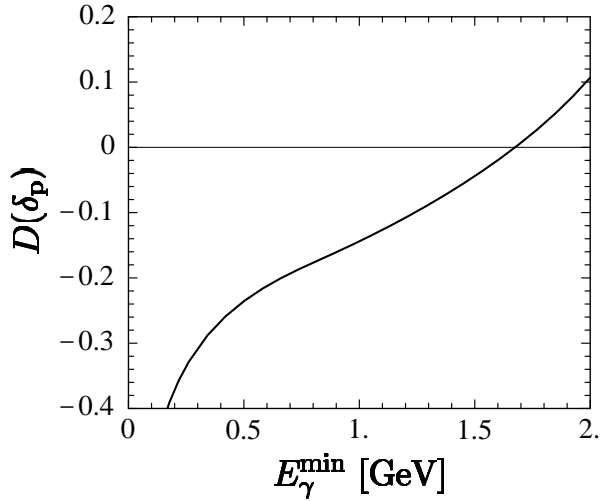}}
\centerline{\parbox{14.5cm}{\caption{\label{fig:Dfun}\small\sl
Theoretical prediction for the function $D(\delta_{\rm p})$}}}
\end{figure}

In Figure~\ref{fig:Dfun}, the function $D(\delta_{\rm p})$ with
$\delta_{\rm p}=1-2 E_\gamma^{\rm min}/m_b$ is shown over a wide range
of cutoff values. For all realistic values of $E_\gamma^{\rm min}$ this
function takes very small values. The corresponding contribution to
$\langle E_\gamma\rangle$ in (\ref{Egamavg}) is typically less than 1\%
and thus negligible. Therefore, to an excellent approximation the
average photon energy in $B\to X_s\gamma$ decays measures the product
of the $b$-quark pole mass times the perturbative series
$C_E[\alpha_s(m_b)]$, which we have computed to $O(\alpha_s)$:
\begin{equation}
   \langle E_\gamma\rangle \approx \frac{m_b}{2}\,C_E[\alpha_s(m_b)]
   \equiv \frac 12\,m_b^{(E)} \,.
\label{mEdef}
\end{equation}
It is well-known that the pole mass of a heavy quark is an
infrared-sensitive quantity, which cannot be unambiguously defined
beyond perturbation theory \cite{BBren,Bigiren}. Formally, this
property appears as a factorial divergence of the expansion
coefficients (i.e.\ an infrared renormalon) in any perturbative series
that relates the pole mass to a short-distance mass, such as the
running mass $\overline{m}_b(m_b)$ in the $\overline{{\sc ms}}$
subtraction scheme. On the other hand, it is clear that the average
photon energy in radiative $B$-meson decays is a physical observable
and as such does not suffer from any ambiguity. Therefore, the quantity
$m_b^{(E)}$ defined in (\ref{mEdef}) has a short-distance nature.
Indeed, it can be shown that the renormalon in the pole mass is exactly
cancelled by a renormalon (i.e.\ a factorial divergence) in the
perturbative series $C_E[\alpha_s]$ \cite{MN}. In other words, the
short-distance mass $m_b^{(E)}$ can be related to any other
short-distance mass without encountering large perturbative
coefficients. In particular, we note that
\begin{equation}
   m_b^{(E)} = \overline{m}_b(m_b) \left( 1 + \frac{49}{54}\,
   \frac{\alpha_s(m_b)}{\pi} + \dots \right) \,.
\end{equation}
Thus, in principle an accurate measurement of the photon spectrum in
$B\to X_s\gamma$ decays would provide for a theoretically clean
determination of the $b$-quark mass. In practice, this determination
will probably be limited by experiment and may not be competitive with
precision determinations of $m_b$ from the analysis of the $\Upsilon$
spectrum.

\section{Hadronic mass distribution}
\label{sec:mass}

\begin{figure}
\epsfxsize=14cm
\centerline{\epsffile{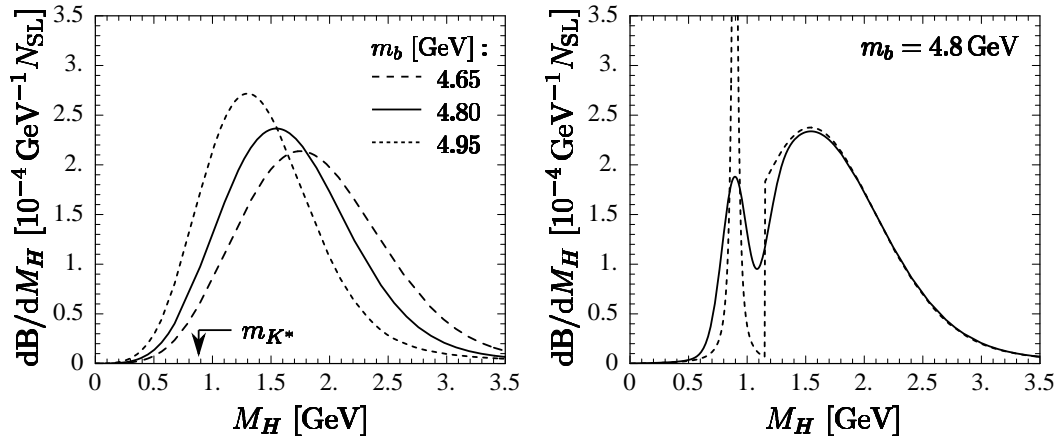}}
\centerline{\parbox{14.5cm}{\caption{\label{fig:mass}\small\sl
Theoretical predictions for the invariant hadronic mass spectrum for
different parameters of the shape function}}}
\end{figure}

In $B\to X_s\gamma$ decays, the invariant mass of the hadronic final
state is related with the photon energy in the $B$ rest frame through
$M_H^2=m_B^2-2m_B E_\gamma$. Therefore, our theoretical results for the
photon spectrum can be translated into predictions for the hadronic
mass spectrum. Since experimentally the measurements of the
photon energy and hadronic mass spectra are quite different, it may be
useful to discuss our results also in terms of the variable $M_H$. In
the left-hand plot in Figure~\ref{fig:mass}, we show the corresponding
spectra obtained using our standard choices for the parameters of the
shape function.

At this point, it may be worthwhile to recall that the theoretical
predictions for the photon energy and hadronic mass spectra must be
understood in the sense of quark--hadron duality. In particular, the
true hadronic mass spectrum in the low-mass region may have resonance
structures due to low-lying kaon states, and thus may look rather
different from our theoretical predictions. To discuss in more detail
how quark--hadron duality works in the present case we distinguish two
kinematic regions: the ``endpoint region'' and the ``resonance
region''. The endpoint region of the photon energy spectrum is
characterized by the condition that $E_\gamma^{\rm
max}-E_\gamma=O(\bar\Lambda)$, where $\bar\Lambda=m_B-m_b$. It is in
this region that the effects of Fermi motion are relevant and determine
the shape of the spectrum. In the endpoint region, the invariant mass
of the hadronic final state is of order $m_B\bar\Lambda\gg\Lambda_{\rm
QCD}^2$, implying that a large number of final states are kinematically
accessible. Under such circumstances, local quark--hadron duality
ensures that the photon and hadronic mass spectra are similar to the
corresponding inclusive spectra predicted by the heavy-quark expansion
even without applying a smearing procedure. In the resonance region, on
the other hand, the invariant mass of the hadronic final state is of
order $\Lambda_{\rm QCD}^2$, implying that the photon energy is very
close to the kinematic endpoint: $E_\gamma^{\rm max}-E_\gamma=
O(\Lambda_{\rm QCD}^2/m_B)$. The heavy-quark expansion does not allow
us to make model-independent predictions for the structure of the
individual resonance contributions. Global quark--hadron duality can,
however, be restored by averaging the spectra over a sufficiently wide
energy interval, whose size is determined by the average level spacing
between the resonance states \cite{PQW}. We will see below that in the
present case the smearing should be done over an interval $\Delta
M_H^2\approx 2$\,GeV$^2$, corresponding to an energy interval $\Delta
E_\gamma\approx 0.2$\,GeV. Note that in the case of the CLEO data such
an averaging is automatically provided by the Doppler shift of the
spectrum due to the motion of the $B$ mesons produced at the
$\Upsilon(4s)$ resonance, and thus the photon spectrum is expected to
be dual to the theoretical spectrum over the entire energy range.

\begin{table}
\centerline{\parbox{14cm}{
\caption{\label{tab:resonances}\small\sl
Mean masses and widths of the lowest-lying hadronic states accessible
in $B\to X_s\gamma$ decays \protect\cite{PDG96}, and the corresponding
photon energies (errors refer to changing $M_H$ by $\pm\Gamma_H$)}}}
\begin{center}
\begin{tabular}{|lccc|}
\hline
State $H$ & $M_H$~[GeV] & $\Gamma_H$~[MeV] & $E_\gamma$~[GeV] \\
\hline\hline
$K\,(n\pi)$ & $\ge 0.629$ & continuum & $\le 2.60$ \\
$K^*(892)$ & 0.894 & ~50 & $2.56\pm 0.01$ \\
$K_1(1270)$ & 1.273 & ~90 & $2.49\pm 0.02$ \\
$K_1(1400)$ & 1.402 & 174 & $2.45\pm 0.05$ \\
$K^*(1410)$ & 1.412 & 227 & $2.45\pm 0.06$ \\
$K_2^*(1430)$ & 1.428 & 103 & $2.45\pm 0.03$ \\
$K_2(1580)$ & 1.580 & 110 & $2.40\pm 0.03$ \\
$K_1(1650)$ & 1.650 & 150 & $2.38\pm 0.05$ \\
$K^*(1680)$ & 1.714 & 323 & $2.36\pm 0.10$ \\
$K_2(1770)$ & 1.773 & 186 & $2.34\pm 0.06$ \\
\hline
\end{tabular}
\end{center}
\end{table}

To make these statements more precise, consider the properties of the
lowest-lying kaon states contributing to $B\to X_s\gamma$ decays, which
are collected in Table~\ref{tab:resonances}. There are six resonances
plus a continuum contribution feeding the photon spectrum in the energy
interval between 2.4 and 2.6\,GeV. Hence, an average over this interval
should be calculable using global quark--hadron duality, although a
much finer resolution cannot be obtained. In the hadronic mass
spectrum, the $K^*(892)$ peak is clearly separated from the rest;
however, the next resonances already have widths exceeding the level
spacing and hence are overlapping. Therefore, we expect that local
duality allows us to predict the hadronic mass spectrum in the region
$M_H\gsim 1.5$\,GeV. Indeed, the pattern of resonances exhibited in
Table~\ref{tab:resonances} suggests a simple but realistic model for
the hadronic mass spectrum consisting of a single Breit--Wigner peak
for the $K^*(892)$ followed by a continuum above a threshold $M_{\rm
cont}$, which is dual to the higher resonance contributions and given
by the inclusive spectrum calculated using the heavy-quark expansion.
This gives
\begin{equation}
   \frac{\rm dB}{{\rm d}M_H}
   = \frac{2 M_H N_{K^*}\,{\rm B}(B\to K^*\gamma)}
          {(M_H^2-m_{K^*}^2)^2 + m_{K^*}^2\Gamma_{K^*}^2}\,
    + \Theta(M_H-M_{\rm cont})\,
   \frac{\rm dB_{\rm incl}}{{\rm d}M_H} \,,
\label{2terms}
\end{equation}
where
\begin{equation}
   N_{K^*} = \frac{m_{K^*}\Gamma_{K^*}}{\displaystyle
    \arctan\!\left( \frac{m_{K^*}}{\Gamma_{K^*}} \right)
    + \frac{\pi}{2}}
\end{equation}
is the normalization of the Breit--Wigner distribution. The exclusive
branching ratio for the decay $B\to K^*\gamma$ can either be taken from
an independent measurement or determined from a fit to the spectrum
itself. The continuum threshold $M_{\rm cont}$ is then fixed by the
requirement that the total branching ratio be the same as that
predicted by the heavy-quark expansion, yielding the condition
\begin{equation}
   \int\limits_0^{M_{\rm cont}}\!{\rm d}M_H\,
   \frac{\rm dB_{\rm incl}}{{\rm d}M_H}
   = \mbox{B}(B\to X_s\gamma)\Big|_{E_\gamma>E_{\rm cont}}
   \stackrel{!}{=} {\rm B}(B\to K^*\gamma) \,,
\label{Econt}
\end{equation}
where $E_{\rm cont}=\frac12(m_B^2-M_{\rm cont}^2)/m_B$. In order to
reduce systematic errors, it will in practice be advantageous to
normalize both sides of (\ref{Econt}) to the total $B\to X_s\gamma$
branching ratio.

To illustrate this method we consider the central value $m_b=4.8$\,GeV
and take from experiment the exclusive branching ratio ${\rm B}(B\to
K^*\gamma)=(0.45\pm 0.17)\times 10^{-4}$ \cite{CLEOexcl} normalized to
our fit result for the total branching ratio given in (\ref{newCLEO2}).
This yields ${\rm B}(B\to K^*\gamma)/{\rm B}(B\to X_s\gamma)=0.17\pm
0.08$. The value of the continuum threshold which reproduces the
central value of this ratio is $M_{\rm cont}\approx 1.15$\,GeV, which
is close to the position of the second resonance $K_1(1270)$. The
dashed line in the right-hand plot in Figure~\ref{fig:mass} shows the
hadronic mass spectrum obtained from (\ref{2terms}) using these
parameters. To smoothen out the sharp structures, we take a convolution
of this curve with a Gaussian smearing function of width
$\sigma_{M_H}=100$\,MeV, which resembles the binning and resolution of
a realistic experiment. The result is shown by the solid curve, which
exhibits a two-peak structure: a narrow peak located at the $K^*$ mass,
whose width is determined by the mass resolution of the experiment, is
followed by a broad bump containing a large number of overlapping
resonances, whose sum is dual to the inclusive spectrum predicted by
the heavy-quark expansion. The position of the second peak is
determined by the $b$-quark mass through $M_H^{\rm bump}\approx
(m_B\bar\Lambda)^{1/2}$, which is located at 1.6\,GeV in the present
case. This relation is just a reflection of the fact that the $b$-quark
mass determines the average photon energy in $B\to X_s\gamma$ decays
(see Section~\ref{sec:spectrum}). We note that such a two-peak
structure is indeed seen in the CLEO data on the hadronic mass
distribution \cite{CLEO}; however, in view of the large experimental
uncertainties it is premature to perform a detailed comparison with the
data.

\section{Impact of New Physics and constraints on extensions of the
Standard Model}
\label{sec:NP}

Measurements of the $B\to X_s\gamma$ branching ratio have been used
extensively to put constraints on the parameters of various extensions
of the Standard Model, such as multi-Higgs models, supersymmetry, or
left--right symmetric models (see Refs.~\cite{Gronau,Hewett} for recent
reviews). In many of these extensions, the Wilson coefficients of the
dipole operators $O_7$ and $O_8$ in the effective weak Hamiltonian
receive additional contributions from new flavour physics at a high
energy scale. The CLEO measurement of the $B\to X_s\gamma$ branching
ratio has been used to extract the magnitudes of the Wilson
coefficients $C_7(m_W)$ and $C_8(m_W)$ and to compare the results with
various model predictions (see, e.g., Refs.~\cite{AGM,HeWe}). Likewise,
new dipole operators of non-standard chirality may be induced. In
general, such New Physics contributions would enter our analysis
through non-standard values of the parameters $\xi_i^{(R)}$ defined in
(\ref{xiLdef}) and (\ref{xiRdef}). If these parameters carry
non-trivial new weak phases, there is potential for observing a large
direct CP asymmetry in the decays $B\to X_s\gamma$, which would be a
striking signature for New Physics \cite{Alex98}. Here, we shall
discuss the impact of New Physics on the CP-average $B\to X_s\gamma$
branching ratio.

The theoretical predictions for the branching ratio and photon spectrum
depend on five real combinations of the parameters $\xi_i^{(R)}$, as
shown in (\ref{xidecomp}) and (\ref{Sdecomp}). It is convenient to
introduce the ratio of New Physics contributions to the
chromo-magnetic and magnetic dipole operators as
\begin{equation}
   \chi = \frac{\xi_8-1}{\xi_7-1} \,.
\end{equation}
A specific New Physics scenario will make a prediction for this
quantity. In models with dipole operators of non-standard chirality, we
assume that $\xi_8^R/\xi_7^R$ is given by the same ratio $\chi$.
Moreover, to simplify the discussion we shall assume that $\chi$ is a
real parameter. This is a good approximation whenever there is a single
dominant New Physics contribution, such as the virtual exchange of a
new heavy particle, contributing to both the magnetic and the
chromo-magnetic dipole operators \cite{Kaga}. With this assumption,
there are only two independent structures appearing in (\ref{xidecomp})
and (\ref{Sdecomp}), which can be taken to be $\mbox{Re}(\xi_7)$ and
$|\xi_7|^2+|\xi_7^R|^2$. Note that the imaginary part of $\xi_7$ enters
only in combination with the right-handed coupling $|\xi_7^R|$,
implying that by measuring the total branching ratio alone one will not
be able to put constraints on the weak phase of $\xi_7$.

As we saw in Section~\ref{sec:spectrum}, New Physics contributions are
unlikely to alter the spectral shape of the photon spectrum in the
experimentally accessible region. We therefore focus our discussion on
the total branching ratio and define the ratio
\begin{equation}
   R_\gamma
   = \frac{{\rm B}(B\to X_s\gamma)}{{\rm B}_{\rm SM}(B\to X_s\gamma)}
   = \frac{K_{\rm NLO}(\delta)}{K_{\rm NLO}^{\rm SM}(\delta)} \,,
\end{equation}
which directly measures the deviation from the Standard Model. From our
result (\ref{newCLEO2}) it follows that at the level of two standard
deviations $0.37<R_\gamma<1.25$. It will be convenient to define a
similar ratio of branching ratios for the rare hadronic decays $B\to
X_{sg}$, which are induced by the chromo-magnetic dipole operator, so
that
\begin{equation}
   R_g = \frac{{\rm B}(B\to X_{sg})}{{\rm B}_{\rm SM}(B\to X_{sg})}
   = \frac{|C_8(m_b)|^2+|C_8^R(m_b)|^2}{|C_8^{\rm SM}(m_b)|^2} \,.
\end{equation}
Whereas the Standard Model predicts the very small value ${\rm B}_{\rm
SM}(B\to X_{sg})\approx 0.2\%$, a much larger branching ratio is
attainable in models with enhanced $b\to s g$ transitions
\cite{Kaga}--\cite{CGG95}. This would increase the production of
charmless final states in hadronic $B$ decays, which could help to
explain the low experimental values of the semileptonic branching ratio
and charm yield. Although the systematic errors in the measurements of
these quantities are large, the results favour values of $\mbox{B}(B\to
X_{sg})$ of order 10\% \cite{Raths,KagaSanBarb}. On the other hand, the
CLEO Collaboration has recently presented a preliminary upper limit on
$\mbox{B}(B\to X_{sg})$ of 6.8\% (90\% CL) \cite{Thorn}. The limit is
increased to 9.0\% if more recent charmed baryon and charmonium yields
are used \cite{KagaSanBarb}. In our graphical analysis below we will
present 5\% and 10\% contours for this branching ratio. Which of these
two is considered to be a more realistic upper bound is left to the
taste of the reader.

The theoretical predictions for the two ratios $R$ can be written as
\begin{eqnarray}
   R_\gamma &=& 1 + A_1(\chi)\,\Big[ \mbox{Re}(\xi_7)-1 \Big]
   + A_2(\chi) \left( |\xi_7|^2+|\xi_7^R|^2 - 1 \right) \,,
   \nonumber\\
   R_g &=& 1 + A_3(\chi)\,\Big[ \mbox{Re}(\xi_7)-1 \Big]
    + A_4(\chi) \left( |\xi_7|^2+|\xi_7^R|^2 - 1 \right) \,,
\end{eqnarray}
where
\begin{eqnarray}
   A_1(\chi)
   &=& \frac{B_{27}+\chi B_{28}+(1-\chi)B_{78}+2\chi(1-\chi)B_{88}}
        {B_{22}+B_{27}+B_{77}+B_{28}+B_{78}+B_{88}}
    \approx 0.46 + 0.020\chi - 0.0027\chi^2 \,, \nonumber\\
   A_2(\chi)
   &=& \frac{B_{77}+\chi B_{78}+\chi^2B_{88}}
            {B_{22}+B_{27}+B_{77}+B_{28}+B_{78}+B_{88}}
    \approx 0.11 + 0.025\chi + 0.0013\chi^2 \,, \nonumber\\
   A_3(\chi) &=& \frac{2\chi(1-\chi+r)}{(1+r)^2}
    \approx 0.43\chi(1-\chi) + 0.50\chi \,, \nonumber\\
   A_4(\chi) &=& \frac{\chi^2}{(1+r)^2} \approx 0.21\chi^2 \,,
\end{eqnarray}
with $r=\sum_{i=1}^8\bar h_i\,\eta^{a_i-14/23}/C_8^{(0)}(m_W)\approx
1.16$. The expressions for the functions $A_1$ and $A_2$ follow
directly from (\ref{xidecomp}), whereas those for $A_3$ and $A_4$
follow from the result for $C_8^{(0)}(\mu_b)$ given in (\ref{evol}).
Since these functions depend on ratios of coefficient functions, their
numerical values are rather stable against variations of the input
parameters. The numerical results quoted above are obtained by taking
central values of all parameters, and in the case of $A_1$ and $A_2$
using $E_\gamma^{\rm min}=1.95$\,GeV for the cutoff on the photon
energy, corresponding to the energy interval covered by the data shown
in Table~\ref{tab:CLEOdata} and used to derive the result
(\ref{newCLEO2}).

\begin{table}
\centerline{\parbox{14cm}{
\caption{\label{tab:xi}\small\sl
Approximate ranges of $\chi$ values for various New Physics
contributions to $C_7$ and $C_8$, characterized by the particles in
penguin diagrams}}}
\begin{center}
\begin{tabular}{|lr|}
\hline
New Physics penguins & $\chi$ \\
\hline\hline
gluino--squark & 2--20 \\
techniscalar & 9 \\
charged Higgs--top & 1.5--2.4 \\
Higgsino--stop & 0.2--2.2 \\
left--right $W$--top & 0.9 \\
scalar diquark--top & $-(0.8\mbox{--}1.3)$ \\
neutral scalar--vectorlike quark & $-8$ \\
\hline
\end{tabular}
\end{center}
\end{table}

Approximate ranges of $\chi$ for several illustrative New Physics
scenarios with $b\to s\gamma$ and $b\to sg$ penguin diagrams containing
new particles in the loop have been discussed in
Ref.~\cite{Alex98}\footnote{In the notation of this reference
$\chi=\xi_{\rm SM}/\xi$ with $\xi_{\rm SM}=-3 C_7^{\rm SM}(m_W)/
C_8^{\rm SM}(m_W)\approx -6.05$.}
and are collected in Table~\ref{tab:xi}. Our aim here is not to carry
out a detailed study of each model, but to give the reader an idea of
the sizable variation that is possible in $\chi$. In supersymmetric
theories, e.g., penguin diagrams with gluino--squark loops imply a
positive value of $\chi$, which can be tuned over a large range by
adjusting the mass ratio $m_{\tilde g}/m_{\tilde q}$. A detailed
analysis of the $B\to X_s\gamma$ branching ratio in this scenario is
presented in Ref.~\cite{CGG95}. Another example with large positive
$\chi$ is provided by models with techniscalars of charge $\frac 16$
\cite{Kaga,JHU,Dobrescu}, whereas an example with large negative $\chi$
is provided by penguin diagrams with new neutral scalars and
vector-like quarks of charge $-\frac 13$. In general, models with large
$|\chi|$ offer the possibility of having a direct CP asymmetry in $B\to
X_s\gamma$ decays of as much as 10--50\% in a large region of model
parameter space \cite{Alex98}. Examples of scenarios which lead to more
moderate $\chi$ values include left--right symmetric models and
multi-Higgs models.

\begin{figure}
\epsfxsize=14cm
\centerline{\epsffile{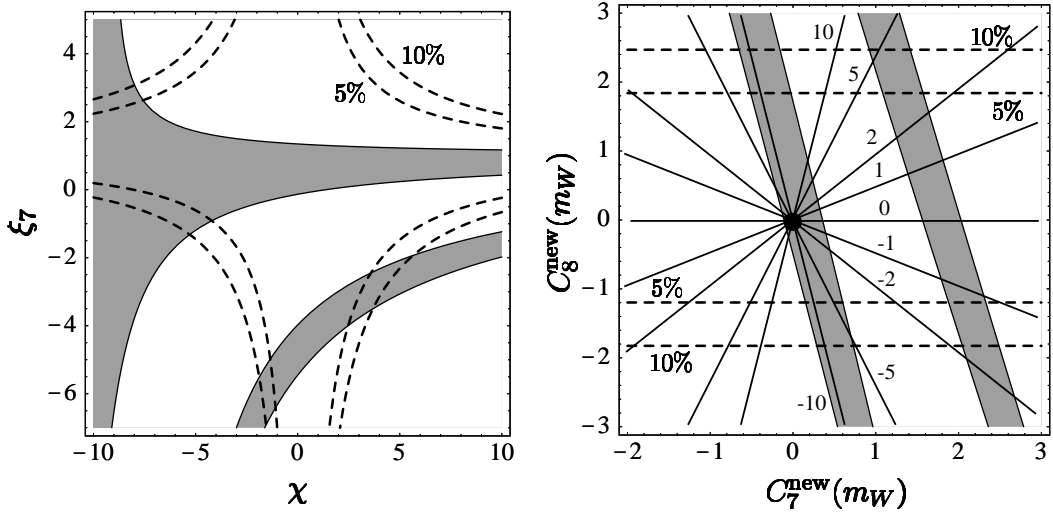}}
\centerline{\parbox{14.5cm}{\caption{\label{fig:NP1}\small\sl
Regions in parameter space allowed by the experimental constraint on
the $B\to X_s\gamma$ branching ratio (bands), and contours for the
$B\to X_{sg}$ branching ratio (dashed lines), assuming
$\mbox{Im}(\xi_7)=\xi_7^R=0$. The thin lines in the right-hand plot
show contours of constant $\chi$ values as indicated, the black circle
corresponds to the Standard Model.}}}
\end{figure}

Let us first assume that the New Physics contributions to $C_7$ and
$C_8$ do not contain new weak phases, and that there are no new
operators with non-standard chirality, i.e.\
$\mbox{Im}(\xi_7)=\xi_7^R=0$. In the left-hand plot in
Figure~\ref{fig:NP1}, we show as a function of $\chi$ the allowed
ranges for $\xi_7$ which satisfy the condition that
$0.37<R_\gamma<1.25$. We also show as dashed lines the 5\% and 10\%
contours for the $B\to X_{sg}$ branching ratio. There are three
different regions to distinguish: (a) for $\chi>3$ only positive values
$\mbox{Re}(\xi_7)\lsim 1$ are allowed, which are close to the Standard
Model value $\xi_7=1$; (b) for $-1<\chi<3$ a second branch of large
negative values of $\mbox{Re}(\xi_7)$ is allowed, which have magnitude
several times larger than in the Standard Model; (c) for $\chi<-1$ only
the first branch remains, and the combined constraints from the $B\to
X_s\gamma$ and $B\to X_{sg}$ branching ratios imply that
$-1<\xi_7<2.5$. The right-hand plot in Figure~\ref{fig:NP1} shows the
same information in a different way, namely in the plane spanned by the
(real) New Physics contributions to the Wilson coefficients $C_7(m_W)$
and $C_8(m_W)$. This figure generalizes the corresponding leading-order
results discussed in Refs.~\cite{AGM,HeWe,CGG95}. It is evident that
for a given New Physics scenario, i.e.\ for a chosen value of $\chi$,
the constraints imposed on the Wilson coefficients are quite
non-trivial. An example is provided by the ``minimal supergravity
model'' \cite{Kane} investigated by Hewett and Wells \cite{HeWe}, who
perform a scan in the SUSY parameter space finding that
$-2.5<\xi_7<5.5$, and $\chi\approx 1$ for choices of parameters
yielding sizable New Physics contributions to $C_7$ and $C_8$. From
Figure~\ref{fig:NP1} it follows that in this model the constraint from
the $B\to X_s\gamma$ branching ratio implies $-0.1<\xi_7<1.2$, which
excludes a considerable fraction of the SUSY parameter space, whereas
there is no constraint from the experimental bound on the $B\to X_{sg}$
branching ratio. Hence, in the minimal supergravity model the Wilson
coefficients must take values close to those predicted by the Standard
Model or somewhat smaller.

\begin{figure}
\epsfxsize=15cm
\centerline{\epsffile{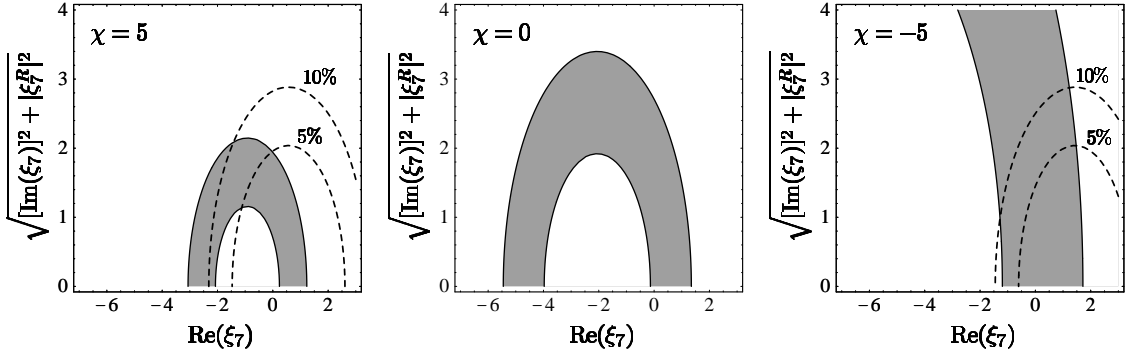}}
\centerline{\parbox{14.5cm}{\caption{\label{fig:NP2}\small\sl
Regions in parameter space allowed by the experimental constraint on
the $B\to X_s\gamma$ branching ratio (bands), and contours for the
$B\to X_{sg}$ branching ratio (dashed lines), in the general case with
non-trivial weak phases and/or non-standard chirality operators}}}
\end{figure}

The situation becomes more complicated if one allows for complex values
of the Wilson coefficients arising from new weak phases (i.e.\
$\mbox{Im}(\xi_7)\ne 0$), or considers the possibility of having dipole
operators with non-standard chirality (i.e.\ $\xi_7^R\ne 0$)
\cite{Rizzo}. Three illustrative cases are shown in
Figure~\ref{fig:NP2}. Note that for large $|\chi|$ the constraint from
the $B\to X_{sg}$ branching ratio is quite non-trivial and puts a
stringent upper bound on the combination
$[\mbox{Im}(\xi_7)]^2+|\xi_7^R|^2$.

\section{Conclusions}
\label{sec:concl}

The inclusive radiative decays $B\to X_s\gamma$ play a key role in
testing the Standard Model and probing the structure of possible New
Physics. We have presented a detailed study of these decays using the
operator product expansion for inclusive decays of heavy hadrons
combined with the twist expansion for the description of decay
distributions near phase-space boundaries. We have updated (and
corrected) the existing next-to-leading order analyses of the total
$B\to X_s\gamma$ branching ratio and added several improvements
concerning the treatment of QED corrections, the analysis of the
renormalization-scale dependence, and the discussion of the sensitivity
to New Physics. In particular, we have argued that the truncation error
of the perturbative expansion in $\alpha_s$ has been underestimated by
previous authors by at least a factor of 2.

Our main focus, however, was to implement a consistent treatment of
bound-state effects related to the soft interactions of the $b$-quark
inside the $B$ meson. These effects cause the Fermi motion of the heavy
quark, which is responsible for the characteristic shape of the photon
energy spectrum in $B\to X_s\gamma$ decays. They lead to the main
theoretical uncertainty in the calculation of the inclusive branching
ratio if a restriction to the high-energy part of the photon spectrum
is imposed, as is necessary in realistic experiments. Fermi motion is
naturally incorporated in the heavy-quark expansion by resumming an
infinite set of leading-twist operators into a non-perturbative shape
function. The main theoretical uncertainty in this description lies in
the value of the $b$-quark mass. Other features associated with the
detailed functional form of the shape function play only a minor role,
particularly if a partial integration over the decay distributions is
implied. We have explained how the value of $m_b$ could, in principle,
be extracted from a measurement of the average photon energy in $B\to
X_s\gamma$ decays. For the Standard Model, we have obtained ${\rm
B}(B\to X_s\gamma)=(2.57\pm 0.26_{-0.36}^{+0.31})\times 10^{-4}$ for
the integral over the high-energy part of the photon spectrum with
$E_\gamma^{\rm lab}>2.2$\,GeV, where the first error reflects the
uncertainty in the input parameters, and the second one the uncertainty
in the calculation of Fermi motion. This prediction agrees with the
CLEO measurement of the same quantity within one standard deviation.
{}From a reanalysis of the CLEO data, we have obtained values for the
total branching ratio that are consistent with the Standard Model
prediction of $(3.29\pm 0.33)\times 10^{-4}$. In the future, an effort
should be made to lower the cutoff on the photon energy to a value of
2\,GeV or less, even if this would increase the experimental systematic
errors. The benefit of such a low cutoff would be that the calculation
of the branching ratio becomes insensitive to the effects of Fermi
motion, reducing the theoretical uncertainty to the level of 10\%.

Besides the photon spectrum, we have studied the invariant hadronic
mass distribution in radiative $B$ decays. Investigating the pattern of
individual hadron resonances contributing to the spectrum, we have
argued that a complementarity between the inclusive theoretical
distribution and the true spectrum should set in for $M_H\gsim
1.5$\,GeV. This leads us to a simple description of the hadronic mass
spectrum with only a single parameter, the $B\to K^*\gamma$ branching
ratio, to be determined from experiment.

Finally, we have investigated the sensitivity of the $B\to X_s\gamma$
branching ratio and photon spectrum to New Physics beyond the Standard
Model and set up a formalism to include possible non-standard
contributions in a straightforward way. New Physics contributions enter
our predictions through the values of parameters $\xi_i^{(R)}$, which
are defined in terms of Wilson coefficient functions at the scale
$m_W$. This formalism allows us to account for non-standard
contributions to the magnetic and chromo-magnetic dipole operators, as
well as operators with right-handed light-quark fields. In the context
of a particular model, all that is needed is to perform a matching
calculation determining the Wilson coefficients in the effective
Hamiltonian at the weak scale. We find that, quite generally, New
Physics contributions would not affect the shape of the photon
spectrum, but of course could change the total branching ratio by a
considerable amount. This implies that the analysis of the photon
energy and hadronic mass spectra, which is crucial for the experimental
determination of the total branching ratio, can be performed without
assuming the correctness of the Standard Model. On the other hand, the
total branching ratio will provide a powerful constraint on the
structure of New Physics beyond the Standard Model, as we have
illustrated with some specific examples.

We believe that our work eliminates the remaining elements of model
dependence present in previous studies of the photon spectrum in $B\to
X_s\gamma$ decays and, therefore, provides a firm theoretical basis for
analyses of experimental data on inclusive radiative $B$ decays. We are
confident that in the near future, when measurements of the photon
spectrum with high statistics will be performed, it will be possible to
derive a value for the $B\to X_s\gamma$ branching ratio that is less
model-dependent than existing ones, thus providing one of the most
sensitive tests of the flavour sector of the Standard Model.

\vspace{0.3cm}
{\it Note added:\/}
While this paper was in preparation the CLEO Collaboration presented a
preliminary update of the $B\to X_s\gamma$ branching ratio, yielding
$\mbox{B}(B\to X_s\gamma)=(2.50\pm 0.47\pm 0.39)\times 10^{-4}$
\cite{APS}. This value has been obtained using the original analysis
adopted in Ref.~\cite{CLEO}. We expect that using our improved
theoretical predictions the central value will increase to about
$2.8\times 10^{-4}$, which is close to the prediction of the Standard
Model.

\vspace{0.3cm}
{\it Acknowledgments:\/}
We would like to thank Ahmed Ali, Gerhard Buchalla, Andrzej Buras, Andrzej
Czarnecki, Gian Giudice, Laurent Lellouch, Guido Martinelli and Mikolaj Misiak for
helpful discussions. We are indebted to Steven Glenn for discussing
aspects of the CLEO measurement of the $B\to X_s\gamma$ branching ratio
and providing the data on the photon spectrum displayed in
Table~\ref{tab:CLEOdata}. A.K.~was supported by the United States
Department of Energy under Grant No.\ DE-FG02-84ER40153.

\setcounter{equation}{0}
\renewcommand{\theequation}{A.\arabic{equation}}
\boldmath
\section*{Appendix~A: QED corrections to $C_7(m_b)$}
\unboldmath

QED corrections affect the theoretical prediction for the $B\to
X_s\gamma$ branching ratio in three ways: there are $O(\alpha)$
matching corrections to the Wilson coefficients $C_i(m_W)$ at the weak
scale, there are $O(\alpha)$ contributions to the matrix elements of
the operators in the effective Hamiltonian at the scale $\mu_b$, and
there are $O(\alpha L)$ corrections to the evolution of the operators
from the scale $m_W$ down to the scale $\mu_b$, where
$L=\ln(m_W/\mu_b)$. The latter corrections are logarithmically
enhanced. They can be accounted for by including the $O(\alpha)$
contributions to the anomalous dimension matrix of the operators in the
effective Hamiltonian in the solution of the renormalization-group
equation. On the other hand, a complete calculation of the remaining
$O(\alpha)$ corrections would be extremely cumbersome. Fortunately, it
is likely that these corrections will be smaller than the
next-to-next-to-leading QCD corrections of order $\alpha_s^2$.

When QED corrections are included, as many as twelve operators in the
effective Hamiltonian mix under renormalization. Besides the
current-current operators $O_1$ and $O_2$ and the dipole operators
$O_7$ and $O_8$, these are four QCD penguin and four electroweak
penguin operators. However, it turns out that to a very good
approximation the mixing of $(O_1,O_2,O_7,O_8)$ with the penguin
operators can be neglected. This approximation reproduces the
leading-order results in (\ref{evol}) exactly but for the terms $\sum_i
h_i\eta^{a_i}$ and $\sum_i \bar h_i\eta^{a_i}$ in $C_7^{(0)}$ and
$C_8^{(0)}$. Numerically, $C_7^{(0)}(m_b)$ is reproduced with an
accuracy of $3\times 10^{-3}$, and $C_8^{(0)}(m_b)$ with an accuracy of
$3.7\%$.

Including QED corrections, the one-loop anomalous dimension matrix in
the truncated operator basis is
\begin{equation}
   \mbox{\boldmath$\gamma$\unboldmath}
   = \frac{\alpha_s}{4\pi}\,\mbox{\boldmath$\gamma_0$\unboldmath}
   + \frac{\alpha}{4\pi}\,\mbox{\boldmath$\Gamma_0$\unboldmath} \,,
\end{equation}
where
\begin{equation}
   \mbox{\boldmath$\gamma_0$\unboldmath} = \left(
   \begin{array}{rrrr}
    -2 & 6 & 0 & 3 \\
    6 & -2 & \frac{416}{81} & ~\frac{70}{27} \\
    0 & 0 & \frac{32}{3} & 0 \\
    0 & 0 & -\frac{32}{9} & \frac{28}{3}
   \end{array} \right) \,, \qquad
   \mbox{\boldmath$\Gamma_0$\unboldmath} = \left(
   \begin{array}{rrrr}
    -\frac83 & 0 & -\frac{208}{81} & -\frac{8}{27} \\
    0 & -\frac83 & -\frac{208}{243} & -\frac{116}{81} \\
    0 & 0 & \frac{16}{9} & -\frac83 \\
    0 & 0 & 0 & \frac89
   \end{array} \right) \,.
\end{equation}
We have determined the entries of the QED matrix
{\boldmath$\Gamma_0$\unboldmath} 
using the results of Ciuchini et al. \cite{Ciuchini94} for the 
contributions to each entry of the QCD matrix {\boldmath$\gamma_0$\unboldmath}
taking into account
their color structure and the electric charges of the quark fields
involved.
We collect the
Wilson coefficient functions $C_i^{(0)}$ into a vector $\vec C$ and
write their evolution as
\begin{equation}
   \vec C(\mu) = {\bf U}(\mu,m_W)\,\vec C(m_W) \,,
\end{equation}
where the evolution matrix satisfies the renormalization-group equation
\begin{equation}
   \mu\,\frac{\rm d}{{\rm d}\mu}\,{\bf U}(\mu,m_W)
   = \mbox{\boldmath$\gamma$\unboldmath}^T\,{\bf U}(\mu,m_W) \,.
\end{equation}
To solve this equation, we make the ansatz
\begin{equation}
   {\bf U}(\mu,m_W) = {\bf K}(\mu)\,{\bf U_0}(\mu,m_W)\,
   {\bf K}^{-1}(m_W) \,,
\end{equation}
where ${\bf U_0}(\mu,m_W)$ is the leading-order QCD evolution matrix.
To
first-order in the electro-magnetic coupling $\alpha$ we may write
\begin{equation}
   {\bf K}(\mu) = {\bf 1} + \alpha \left(
   \frac{\bf K_0}{\alpha_s(\mu)} + \dots \right) \,,
\label{Kmu}
\end{equation}
where the ellipses represent terms that do not contribute at
leading-logarithmic order. Inserting this ansatz into the
renormalization-group equation satisfied by ${\bf K}(\mu)$, we obtain
${\bf K_0}$ from the solution of the algebraic equation
\begin{equation}
   2\beta_0 {\bf K_0}
   = \mbox{\boldmath$\Gamma_0$\unboldmath}\hspace{-0.2cm}^T
   + [\mbox{\boldmath$\gamma_0$\unboldmath}\hspace{-0.17cm}^T,
   {\bf K_0}] \,,
\label{algeb}
\end{equation}
where $\beta_0=\frac{23}{3}$ is the first coefficient of the QCD
$\beta$-function for $n_f=5$ light quark flavours. Let us introduce the
matrix ${\bf V}$ that diagonalizes the QCD anomalous dimension matrix,
i.e.\
\begin{equation}
   {\bf V}^{-1}
   \mbox{\boldmath$\gamma_0$\unboldmath}\hspace{-0.17cm}^T\,
   {\bf V} \equiv \mbox{diag}(\gamma_1,\gamma_2,\gamma_3,\gamma_4) \,,
\end{equation}
and denote ${\bf k}={\bf V}^{-1}{\bf K_0}\,{\bf V}$ and ${\bf g}={\bf
V}^{-1}\mbox{\boldmath$\Gamma_0$\unboldmath}\hspace{-0.2cm}^T\,{\bf
V}$. Then the solution of (\ref{algeb}) yields
\begin{equation}
   k_{ij} = \frac{g_{ij}}{2\beta_0-(\gamma_i-\gamma_j)} \,.
\end{equation}
The result for the evolution matrix ${\bf U}(\mu,m_W)$ can now be
written in the form ${\bf U}(\mu,m_W)={\bf V}\,{\bf u}\,{\bf V}^{-1}$,
where
\begin{equation}
   u_{ij} = \eta_i\,\delta_{ij} + \alpha\,k_{ij} \left(
   \frac{\eta_i}{\alpha_s(\mu)} - \frac{\eta_j}{\alpha_s(m_W)}
   \right) \,;\quad
   \eta_i = \left( \frac{\alpha_s(m_W)}{\alpha_s(\mu)}
   \right)^{\gamma_i/2\beta_0} \,.
\end{equation}
This general result has previously been derived by Buchalla et al.\
\cite{Buchalla}. The evaluation of this relation for our particular
case leads to the expression for the QED coefficient $C_7^{({\rm
em})}(\mu_b)$ given in (\ref{fancy}). The coefficients of
$C_7^{(0)}(m_W)$ and $C_8^{(0)}(m_W)$ in this result are exact, whereas
the remaining terms are only approximate because of the truncation of
the operator basis. However, as in the case of the QCD evolution we
expect that from a numerical point of view the truncation of the basis
is justified. Expanding the result (\ref{fancy}) to first order in
$\alpha_s$ we recover the formula of Czarnecki and Marciano
\cite{CzMa}:
\begin{equation}
   \frac{\alpha}{\alpha_s(\mu_b)}\,C_7^{({\rm em})}(\mu_b)
   = \frac{\alpha}{4\pi} \left( \frac{208}{243}
   - \frac{16}{9}\,C_7^{(0)}(m_W) \right) \ln\frac{m_W}{\mu_b}
   + \dots \,.
\label{CzMa}
\end{equation}
Numerically, the resummed expression in (\ref{fancy}) is smaller than
the naive result (\ref{CzMa}) by a factor of about 0.55.

We note that Czarnecki and Marciano also include a particular type of
matching correction to the coefficient $C_7(m_W)$ at the weak scale,
arising from fermion-loop insertions on the $W$-propagator in the
top-quark--$W$ penguin diagram \cite{CzMa}. The resulting contribution
is $\Delta C_7(m_W)\approx 0.53\alpha$, which at the scale $m_b$ yields
a contribution $\Delta C_7(m_b)\approx 0.36\alpha\approx 2.6\times
10^{-3}$. Since there are many other matching corrections that have not
yet been calculated, and since there are similar $O(\alpha)$
contributions to the matrix elements of the local operators $O_i$ that
are neglected, we see no compelling reason to include this particular
matching contribution. Its effect on the total branching ratio does not
exceed the level of 1\% and is thus safely within the theoretical
uncertainty of $\pm 2\%$, which we assign to higher-order electroweak
corrections.

\setcounter{equation}{0}
\renewcommand{\theequation}{B.\arabic{equation}}
\boldmath
\section*{Appendix~B: Photon spectrum in decays of $B$ mesons produced
at the $\Upsilon(4s)$ resonance}
\unboldmath

We denote by $E_\gamma^{\rm lab}$ the photon energy measured in the
laboratory, and by $E_\gamma$ the one measured in the $B$ rest frame.
If ${\rm dB}/{\rm d}E_\gamma$ is the photon spectrum in the $B$ rest
frame, the corresponding spectrum measured in the laboratory is
\begin{equation}
   \frac{\mbox{dB}}{\mbox{d}E_\gamma^{\rm lab}}
   = \frac{1}{\beta_+ - \beta_-}
   \int\limits_{\beta_- E_\gamma^{\rm lab}}^{E_1(E_\gamma^{\rm lab})}\!
   \mbox{d}E_\gamma\,\frac{1}{E_\gamma}\,
   \frac{\mbox{dB}}{\mbox{d}E_\gamma} \,,
\label{LABspec}
\end{equation}
where
\begin{equation}
   \beta_\pm = \sqrt{\frac{1\pm\beta}{1\mp\beta}}\approx 1\pm\beta \,,
   \qquad
   E_1(E_\gamma^{\rm lab}) = \mbox{min}(\beta_+ E_\gamma^{\rm lab},
    E_\gamma^{\rm max}) \,.
\end{equation}
The maximum photon energy in the laboratory is $(E_\gamma^{\rm
lab})_{\rm max} = \beta_+ E_\gamma^{\rm max}$.

It is straightforward to calculate from (\ref{LABspec}) the effect of
the boost on the partially integrated branching ratios. Let us define
the difference
\begin{equation}
   \Delta(E_0) = \int\limits_{E_0}^{(E_\gamma^{\rm lab})_{\rm max}}\!
   \mbox{d}E_\gamma^{\rm lab}\,
   \frac{\mbox{dB}}{\mbox{d}E_\gamma^{\rm lab}}
   - \int\limits_{E_0}^{E_\gamma^{\rm max}}\!
   \mbox{d}E_\gamma\,\frac{\mbox{dB}}{\mbox{d}E_\gamma} \,.
\end{equation}
Then, provided that $E_0<\beta_- E_\gamma^{\rm max}$, we find that
\begin{eqnarray}
   \Delta(E_0) &=& \frac{1-\beta}{2\beta}
    \int\limits_{E_0}^{\beta_+ E_0}\!\mbox{d}E_\gamma\,
    \frac{\mbox{dB}}{\mbox{d}E_\gamma} \left(
    1 - \frac{\beta_+ E_0}{E_\gamma} \right)
    + \frac{1+\beta}{2\beta}
    \int\limits_{\beta_- E_0}^{E_0}\!\mbox{d}E_\gamma\,
    \frac{\mbox{dB}}{\mbox{d}E_\gamma} \left(
    1 - \frac{\beta_- E_0}{E_\gamma} \right) \nonumber\\
   &=& -\frac{\beta^2 E_0^3}{6} \left( \frac{{\rm d}}{{\rm d}E}\,
    \frac{1}{E}\,\frac{{\rm dB}}{{\rm d}E} \right)_{\!E=E_0}
    + O(\beta^4) \,,
\end{eqnarray}
i.e.\ the effect is quadratic in the small quantity $\beta$. With
$E_0=2.2$\,GeV as in the CLEO analysis, we find that
$\Delta(E_0)\approx -(0.04\mbox{--}0.09)\times 10^{-4}$ depending on
the parameters of the shape function, which is a very small effect.

\setcounter{equation}{0}
\renewcommand{\theequation}{C.\arabic{equation}}
\boldmath
\section*{Appendix~C: Results for the functions $s_{ij}(y)$ and
$d_{ij}(\delta)$}
\unboldmath

The functions $s_{ij}(y)$ entering the theoretical expressions for the
photon spectrum are given by the derivatives of the functions
$f_{ij}(\delta)$ in (\ref{fij}) through $s_{ij}(y)=f_{ij}'(1-y)$.
Explicitly, we find
\begin{eqnarray}
   s_{77}(y) &=& \frac{1}{3} \left[ 7+y-2y^2-2(1+y)\ln(1-y) \right]
    \,, \nonumber\\
   s_{88}(y) &=& \frac{1}{27} \left\{ \frac{2(2-2y+y^2)}{y} \left[
    \ln(1-y) + 2\ln\frac{m_b}{m_s} \right]
    - 2y^2 - y - \frac{8(1-y)}{y} \right\} \,,
    \nonumber\\
   s_{78}(y) &=& \frac{8}{9} \left[ \frac{1-y}{y}\ln(1-y) + 1
    + \frac{y^2}{4} \right] \,, \nonumber\\
   s_{22}(y) &=& \frac{16}{27} \int\limits_0^y\!{\rm d}x\,
    (1-x)\,\left|\, \frac{z}{x}\,G\!\left(\frac{x}{z}\right)
    + \frac 12 \,\right|^2 \,, \nonumber\\
   s_{27}(y) &=& -3 s_{28}(y)
    = - \frac{8z}{9} \int\limits_0^y\!{\rm d}x\,\mbox{Re}\!\left[
    G\!\left(\frac{x}{z}\right) + \frac{x}{2z} \right] \,.
\end{eqnarray}

The functions $d_{ij}(\delta)$ entering the expression for the average
photon energy in (\ref{Egamavg}) are given by
\begin{equation}
   d_{ij}(\delta) = \int\limits_0^\delta\!{\rm d}x\,f_{ij}(x)
   - \delta\,f_{ij}(\delta) \,.
\end{equation}
We find
\begin{eqnarray}
   d_{77}(\delta) &=& \frac{2\delta^2}{9}\,(3-\delta)\ln \delta
    - \frac{4\delta^2}{3} - \frac{7\delta^3}{27} + \frac{\delta^4}{6}
    \,, \nonumber\\
   d_{88}(\delta) &=& \frac{8}{27} \left( \ln\frac{m_b}{m_s} - 1
\right)
    \bigg[ \ln(1-\delta) + \delta + \frac{\delta^2}{4}
    + \frac{\delta^3}{6} \bigg] \nonumber\\
   &&\mbox{}+ \frac{4}{27} \left[ \frac{\pi^2}{6} - L_2(1-\delta)
    + \bigg( \delta + \frac{\delta^2}{4} + \frac{\delta^3}{6} \bigg)
    \ln\delta - \delta - \frac{\delta^2}{4} - \frac{5\delta^3}{36}
    + \frac{\delta^4}{8} \right] \,, \nonumber\\
   d_{78}(\delta) &=& \frac89 \left[ \frac{\pi^2}{6} - L_2(1-\delta)
    + \bigg( \delta + \frac{\delta^2}{2} \bigg) \ln\delta
    - \delta - \frac{7\delta^2}{8} + \frac{\delta^3}{6}
    - \frac{\delta^4}{16} \right] \,, \nonumber\\
   d_{22}(\delta) &=& - \frac{8}{27} \int\limits_0^1\!{\rm d}x\,
    (1-x)(1-x_\delta)^2\,\left|\, \frac{z}{x}\,
    G\!\left(\frac{x}{z}\right) + \frac 12 \,\right|^2 \,, \nonumber\\
   d_{27}(\delta) &=& -3 d_{28}(\delta)
    = \frac{4z}{9} \int\limits_0^1\!{\rm d}x\,(1-x_\delta)^2\,
    \mbox{Re}\!\left[ G\!\left(\frac{x}{z}\right) + \frac{x}{2z}
    \right] \,.
\end{eqnarray}
The function $\bar d_{77}(\delta)$ is defined as
\begin{equation}
   \bar d_{77}(\delta) = d_{77}(\delta)
   + \delta \left( 1 + \frac43\ln\delta \right) + \frac{23}{54} \,.
\label{bard77}
\end{equation}
The second term is the contribution of the Sudakov logarithms. It is
this contributions which, in large orders, develops a factorial
divergence that cancels the infrared renormalon of the pole mass
\cite{MN}. The last term in (\ref{bard77}) is adjusted such that $\bar
d_{77}(1)=0$. This condition defines the coefficient $C_E[\alpha_s]$ in
(\ref{Egamavg}).

\end{document}